\documentclass[useAMS,usenatbib,onecolumn]{mn2e}
\usepackage{psfig,subfigure}
\usepackage{graphicx}

\def\bI{{\bf I}}
\def\bw{{\bf w}}
\def\bl{{\bf l}}
\def\bO{{\bf \Omega}}
\def\dfl{{\tilde f}_\bl}
\def\ldo{\bl\cdot\bO}

\def\erf{\mathop{\rm erf}\nolimits}
\def\apj{ApJ}

\title[Time-dependent secular evolution]{Time-dependent secular
  evolution in galaxies}

\author[M. D. Weinberg]{Martin D. Weinberg\thanks{E-mail:
    weinberg@astro.umass.edu}\\
  Department of Astronomy, University of Massachusetts, Amherst,
}

\begin{document}

\label{firstpage}

\date{\today}
\pagerange{\pageref{firstpage}--\pageref{lastpage}} \pubyear{2004}

\maketitle
\begin{abstract}
  \citet[LBK]{Lynden-Bell.Kalnajs:72} presented a useful formula for
  computing the long-range torque between spiral arms and the disk at
  large. The derivation uses second-order perturbation theory and
  assumes that the perturbation slowly grows over a very long time
  (the {\em time-asymptotic limit}).  This formula has been widely
  used to predict the angular momentum transport between spiral arms
  and stellar bars between disks and dark-matter halos.  However, this
  paper shows that the LBK time-asymptotic limit is {\em not}
  appropriate because the characteristic evolution time for galaxies
  is too close to the relevant dynamical times.  We demonstrate that
  transients, not present in the time-asymptotic formula, can play a
  major role in the evolution for realistic astronomical time scales.
  A generalisation for arbitrary time dependence is presented and
  illustrated by the bar--halo and satellite--halo interaction.  The
  natural time dependence in bar-driven halo evolution causes
  quantitative differences in the overall torque and qualitative
  differences in the physical- and phase-space location of angular
  momentum transfer.  The time-dependent theory predicts that four
  principal resonances dominate the torque at different times and
  accurately predicts the results of an N-body simulation.  In
  addition, we show that the Inner Lindblad Resonance (ILR) is
  responsible for the peak angular momentum exchange but, due to the
  time dependence, the changes occur over a broad range of energies,
  radii and frequencies.  We describe the implication of these
  findings for the satellite--halo interaction using a simple model
  and end with a discussion of possible impact on other aspects
  secular galaxy evolution.
\end{abstract}


\section{Introduction}
\label{sec:intro}

A near-equilibrium galaxy only evolves through the excitation of some
non-axisymmetric structure such as spiral arms, bars, infalling
satellites, etc.  The resulting ``waves'' are likely actors in
triggering star formation, mediating inward gas flow thereby fueling
AGN, heating the stellar disk, among other observable phenomena.  In
addition, the overall evolution of near equilibrium galaxies is caused
by the collective transport of energy and angular momentum from these
same structures.  An accurate theory of secular evolution is essential
for developing our understanding of these interrelated dynamics.

The first step is predicting the evolution of the non-axisymmetric
distortions.  The underlying dynamical mechanism for the stellar and
dark matter components is understood.  At the level of individual
orbits, an excitation such as a bar, a spiral pattern, or orbiting
satellite produces a periodic distortion on all orbits, much like a
driven harmonic oscillator.  Although the torque distorts a typical
orbit, this distortion averages to zero over many orbital periods for
most orbits.  However, for nearly closed orbits\footnote{A closed
orbit is one with integral commensurabilities between the orbital
frequencies and the pattern frequency.}, the net torque does not
vanish.  An adiabatic invariant is broken by the vanishing precession
frequency of the closed orbit; clearly, there is no adiabatic
invariant for a degree of freedom with no motion.  The perturbation
then couples to the oscillatory distortion for these orbits and net
torque is transferred to or from these orbits.  This coupling between
the forced oscillation and the perturbation will be second order in
the perturbation strength. In time, the perturbation itself and the
underlying equilibrium is slowly changed by this torque.  This leads
to a finite measure of orbits that transfer angular momentum to and
from the pattern \citep[for additional discussion, see][hereafter
Paper 2]{Weinberg.Katz:04}.

Although the first-order perturbation may dominate the instantaneous
changes to phase space, the second-order changes describe the net
changes that will persist after many dynamical times.  This process
was described mathematically by \citet[hereafter
LBK]{Lynden-Bell.Kalnajs:72} in the limit that the change in the
perturbation is very slow but still fast enough that the orbital
perturbations remain linear.  LBK derived a formula describing the
exchange of angular momentum between a spiral pattern and the rest of
the disk assuming that the perturbation began infinitely long in the
past, the time-asymptotic limit.  In this limit, any vestigial
response from the formation of the pattern is gone.  The LBK approach
has been widely applied to estimate secular evolution
\citep[e.g.][]{Goldreich.Tremaine:79b,Weinberg:85,Zhang:98,Athanassoula:03}.
\cite{Carlberg.Sellwood:85} took the first step towards a
time-dependent generalisation of the LBK formalism for studying
secular disk evolution.  They included time-dependent transients but
evaluated the secular changes in the infinite time limit.  This is
appropriate for studying the cumulative effect of short-lived
transients.  However, the perturbation theory may be fully generalised
to treat arbitrary time-dependent perturbations over arbitrary
intervals of time.  This allows us to treat the long-lived bar and
satellite interactions over the full age of a galaxy with some
surprising results.

In attempting a detailed description of the angular momentum exchange
between a rotating bar and a dark halo in an N-body simulation (see
Paper 2), I found a significant discrepancy between the predictions of
the second-order LBK perturbation theory and the N-body simulation.
This discrepancy persisted for perturbations over a wide range of bar
amplitudes and scaled with time and perturbation amplitude as expected
from second-order linear theory, implying that its source is not a
break down in linearity.  Rather, the problem is due to the assumption
of an infinitely slow growth of the perturbation.  Not only is the
number of characteristic dynamical times in a galactic age modest, the
growth of a bar, arms and of course satellites is most likely a small
fraction of the galactic age.  Because patterns often appear over
several orbital periods and the total number of orbital periods
available are small, transients may be significant and the
time-asymptotic limit does not apply.  I will show in this paper that
a finite-time-limit generalisation of the LBK formula gives
quantitatively and qualitatively different results.  Quantitatively,
we will see that the overall torque is smaller than that computed from
the LBK formula for a rotating bar.  Qualitatively, different
commensurabilities are important at different epochs of the evolution.
Because the location of the resonances in physical space are important
for the long-term evolution of galaxy, including the finite-time
response is important for understanding this evolution.  In this
limit, the time dependence is more than simply a transient but
embodies the galaxy's evolutionary history.

I will present a new, generalised secular evolution formula and describe
the important details of the dynamics in \S\ref{sec:method}.  The new
torque formula replaces the delta function in the LBK formula by an
integral over the time-dependent perturbation and is similarly
straightforward to apply.  Two specific applications are presented in
\S\ref{sec:examples}: bar--dark halo coupling and satellite--dark halo
coupling.  N-body simulations will demonstrate the importance of the
finite-time response to the net torque and illustrate the discrepancy
with the LBK formula.  In \S\ref{sec:barslow}, we will see that the
ILR dominates the bar evolution and use this new formalism to
explicitly predict the location of the angular momentum deposited. The
same dynamics is then applied to sinking satellites in
\S\ref{sec:sinksat}.  We will conclude with \S\ref{sec:summary}.

\section{Basic principles}
\label{sec:method}

The total torque on an ensemble of orbits perturbed from equilibrium is
\begin{equation}
\left\langle{dL_z\over dt}\right\rangle =
\int_\Gamma d^3x d^3v \, f({\bf x}, {\bf
  v}, t) \, {dL_z\over dt}({\bf x}, {\bf v})
\end{equation}
where $\Gamma$ is the phase-space domain and $f({\bf x}, {\bf v}, t)$
is the phase-space distribution function, ${\dot L}_z\equiv dL_z/dt$
is the torque per orbit, and $\langle\cdot\rangle$ denotes the
phase-space average.  The angular momentum $L_z$ is a conserved
quantity in an equilibrium axisymmetric system and therefore ${\dot
L}_z$ is first-order in the perturbation amplitude.  The distribution
function has, of course, non-vanishing zeroth- and first-order terms.
Over a long period of time, the first-order contribution to
$\langle{\dot L}_z\rangle$ averages to zero.  At the next order, the
oscillating components from the first-order dependence in
$\langle{\dot L}_z\rangle$ and the first-order induced change in the
distribution function, $f_1({\bf x}, {\bf v}, t)$, coherently
reinforce each other, leading to a non-vanishing contribution as
$t\rightarrow\infty$.  However, a closed-form solution to this problem
does not demand the $t\rightarrow\infty$ limit.  We will derive the
first-order changes in these two quantities in \S\ref{sec:dlzdt} and
\S\ref{sec:f1} and put these together \S\ref{sec:torque} to derive the
torque for a finite duration perturbation.

\subsection{Change in angular momentum for a single orbit}
\label{sec:dlzdt}

Hamilton's equations determine the rate of change in angular momentum
for each orbit as follows:
\begin{equation}
  {dL_z\over dt} = {\partial L_z\over\partial t} + [H, L_z]
\end{equation}
where $[\cdot,\cdot]$ are the Poisson brackets.  Because $L_z$ is a
conserved quantity in the absence of any perturbation, this equation
becomes
\begin{equation}
  {dL_z\over dt} = {\partial H\over\partial\bI}\cdot{\partial
    L_z\over\partial\bw} - {\partial H\over\partial\bw}\cdot{\partial
    L_z\over\partial\bI} = -{\partial H_1\over\partial\bw}\cdot{\partial
    L_z\over\partial\bI} + {\cal O}(A^2)
    \label{eq:orbtorq}
\end{equation}
using action-angle variables $\bI$, $\bw$, $H_1$ is the first-order,
perturbed Hamiltonian and $A$ is the perturbation amplitude.  The
subscripts `0' and `1' will indicate zeroth- and first-order
quantities, respectively.

The orbits in an axisymmetric equilibrium galaxy are purely
quasi-periodic in the angles $\bw$ and the actions $\bI$ are
invariant.  The advantage of action-angle variables is that any
phase-space quantity may be expanded in a Fourier series in $\bw$.  In
particular, we now expand $H_1$ in action-angle variables:
\begin{equation}
  H_1(\bI, \bw, t) = \sum_\bl H_{1\,\bl}(\bI, t) e^{i\bl\cdot\bw}
  \label{eq:h1expand}
\end{equation}
where
\begin{equation}
  H_{1\,\bl}(\bI, t) = {1\over(2\pi)^3} \int_0^{2\pi} d\bw
  e^{-i\bl\cdot\bw} H_1(\bI, \bw, t)
  \label{eq:ftcoef}
\end{equation}
and $\bl=(l_1, l_2, l_3)$ is a vector of integers, one for every
degree of freedom.  Applying the angle transform to equation yields
(\ref{eq:orbtorq})
\begin{equation}
  {dL_z\over dt} = -\sum_\bl il_3 H_{1\,\bl}(\bI, t) e^{i\bl\cdot\bw}.
  \label{eq:torque1}
\end{equation}
Physically, equations (\ref{eq:orbtorq}) and (\ref{eq:torque1})
describe the first-order evolution of a particular orbit's $z$
angular momentum component.

\subsection{Changes in the distribution function}
\label{sec:f1}

The development begins with the first-order solution of the linearised
collisionless Boltzmann equation (CBE):
\begin{equation}
  {\partial f_1\over\partial t} + {\partial H_0\over\partial\bI}\cdot
      {\partial f\over\partial\bw} - {\partial H_1\over\partial\bw}\cdot{\partial
        f_0\over\partial\bI} = 0.
\end{equation}
The quantities $f_0$ and $H_0$ are the unperturbed phase-space
distribution function and the unperturbed Hamiltonian, etc.  We will
solve for the phase-space dependence by an action-angle expansion (as
in eq. \ref{eq:h1expand}) and for the time evolution by Laplace
transform of the CBE.  We denote a Laplace transformed variable by a
tilde and an action-angle transformed variable by subscript $\bl$.
The Fourier-Laplace transform of the CBE is
\begin{equation}
  s\dfl + i\ldo\dfl - i\bl\cdot{\partial f_o\over\partial{\bf
  I}}{{\tilde H}_{1\,\bl}} = 0
\label{eq:lcbe}
\end{equation}
or, upon solving for $\dfl$,
\begin{equation}
  \dfl(\bI, s) = 
  {i\bl\cdot{\partial f_o\over\partial\bI} \over s + i\ldo}
  e^{i\bl\cdot{\bf w}} {\tilde H}_{1\,\bl}(\bI, s).
  \label{eq:dfser2}
\end{equation}
where $\bO\equiv\partial H_o/\partial\bI$.  We will now show that the
Laplace transform rather than inclusion of a term $\exp(\epsilon t)$
the limit $\epsilon\rightarrow0$ to impose the arrow of time allows
one to solve for an arbitrary time dependence.  The total perturbing
potential $H_1$ is the response of the galaxy combined with the
external perturbation.  The quantity ${\tilde H}_{1\,\bl}(\bI, s)$ is
the Laplace transform of the time-dependent coefficient of the
action-angle expansion describing the perturbation:
\begin{equation}
  {\tilde H}_{1\,\bl}(\bI, s) \equiv \int^\infty_0 dt e^{-st} 
  H_{1\,\bl}(\bI, t).
  \label{eq:h1laplace}
\end{equation}
In cases where the perturbation is conveniently described in spherical
harmonics, intrinsic symmetries allow ${\tilde H}_{1\,\bl}$ to be
further simplified \citep[see][]{Tremaine.Weinberg:84*2}.  

The desired solution is the inverse Laplace transform of the series:
\begin{equation}
  f_1(s) = \sum_{l_1,l_2,l_3=-\infty}^{\infty} \dfl(\bI, s)
  e^{i(l_1 w_1 + l_2 w_2 + l_3 w_3)}
  \label{eq:dfser}
\end{equation}
where each coefficient $\dfl$ is given by equation (\ref{eq:dfser2}).
The inverse Fourier-Laplace transform of equation (\ref{eq:dfser}) is
straightforward albeit cumbersome.  Recall that the product of the
first-order distribution function and the first-order torque per orbit
gives the second-order torque.  To proceed we first need to transform
equation (\ref{eq:dfser2}) back to the time domain.  Combining
equations (\ref{eq:dfser}) and (\ref{eq:h1laplace}), the inverse
Laplace transform of equation (\ref{eq:dfser2}) requires evaluation of
the following expression:
\begin{eqnarray}
  {1\over2\pi i}\int^{c+i\infty}_{c-i\infty} ds {e^{st}\over s+i\ldo}
  \int^\infty_0 dt^\prime e^{-st^\prime} H_{1\,\bl}(\bI, t^\prime)
  &=&
  \int^\infty_0 dt^\prime H_{1\,\bl}(\bI, t^\prime)
  {1\over2\pi i} \int^{c+i\infty}_{c-i\infty} ds 
  {e^{s(t-t^\prime)}\over s+i\ldo}
  \nonumber \\
  &=&
  \int^\infty_0 dt^\prime H_{1\,\bl}(\bI, t^\prime)
  \cases{
    e^{-i\ldo(t-t^\prime)} & if $t>t^\prime$\cr
    0 & otherwise\cr
  }
  \nonumber \\
  &=&
  \int^t_0 dt^\prime H_{1\,\bl}(\bI, t^\prime) e^{i\ldo(t^\prime-t)}
\end{eqnarray}
where $c$ is more positive than the any exponent of divergence in
$H_{1\,\bl}(\bI, t)$. This is not restrictive for any realistic
astronomical perturbation. The second equality follows by exchanging
the order of integration and using Cauchy's theorem.  Putting this
together, we have the following time-dependent solution for the
first-order distribution function
\begin{equation}
  f_1(\bw, \bI, t) = \sum_{l_1,l_2,l_3=-\infty}^{\infty}
  i\bl\cdot{\partial f_o\over\partial\bI}
  e^{i\bl\cdot{\bf w}} 
  \left\{\int^t_0dt^\prime H_{1\,\bl}(\bI, t^\prime) e^{i\ldo (t^\prime -t)}\right\}
  \label{eq:dfser3}
\end{equation}

\subsection{Second-order torque}
\label{sec:torque}

To get the second-order torque as a function of actions (or other
conserved quantities following Jeans' theorem), we multiply by
equation (\ref{eq:torque1}) by equation (\ref{eq:dfser3}) and integrate
over all of phase space:
\[
\left\langle {dL_z\over dt}\right\rangle \equiv
\int d\bI\int d\bw \, {dL_z\over dt} \left[f_o(\bI) + f_1(\bI, \bw) +
  \cdots\right] = 
\int d\bI\int d\bw \, {dL_z\over dt} f_1(\bI, \bw) + {\cal O}(A^3).
\]
The first term is oscillatory in $\bw$ and vanishes.  
The product of the two first-order solutions, denoted here by
subscripts $\bl$ and $\bl^\prime$, has an angle dependence of the
form
\[
\exp(-i\bl^\prime\cdot\bw) \exp(i\bl\cdot\bw) = \exp(i[\bl -
\bl^\prime]\cdot\bw)
\]
and therefore the angle integral is only non-zero when
$\bl^\prime=\bl$.  Integrating this product over the remaining
phase-space variables (actions) gives an expression for the total
torque on the phase space from the perturbation:
\begin{eqnarray}
  \left\langle {dL_z\over dt}\right\rangle&=& -(2\pi)^3
  \sum_{l_1,l_2,l_3=-\infty}^{\infty} l_3 \int d\bI\,\bl\cdot{\partial
  f_o\over\partial\bI} \left\{\int^t_0dt^\prime H_{1\,\bl}(\bI,
  t^\prime) e^{i\ldo (t^\prime-t)}\right\} H_{1\,-\bl}(\bI, t).
  \label{eq:dfser4}
\end{eqnarray}
Since physical quantities are real, equation (\ref{eq:h1expand})
implies $H_{1\,\bl}^\ast(\bI, t) = H_{1\,-\bl}(\bI, t)$.  We may now
change variables from actions to energy and angular momentum and the
third action $L_z$, the $z$-component of the angular momentum, in
terms of its ratio to the total angular momentum $J$ as follows:
$\cos\beta\equiv L_z/J$.  The angle $\beta$ describes the colatitude
of the orbital plane.  Equation (\ref{eq:dfser4}) becomes
\begin{equation}
  \left\langle {dL_z\over dt}\right\rangle =
  -(2\pi)^3 \sum_{l_1,l_2,l_3=-\infty}^{\infty} l_3 \int \int {dE dJ J
    d(\cos\beta) \over \Omega_1(E, J)}\ \bl\cdot{\partial
  f_o\over\partial\bI} \left\{\int^t_0 dt^\prime e^{i\ldo
  (t^\prime-t)} H_{1\,\bl}(\bI, t^\prime) \right\}
  H_{1\,\bl}^\ast(\bI, t).
  \label{eq:LBK1}
\end{equation}

For a fixed gravitational potential profile whose only time dependence
is through a rotating pattern, the time dependence in the perturbation
separates as $H_1(\bI,\bw, t) = a(t) H_1(\bI, \bw)$ and therefore
$H_{1\,\bl}(\bI, t) = a(t) H_{1\,\bl}(\bI)$.  Furthermore for a
constant pattern speed, the time-dependent coefficients have the form:
$a(t) = a(0)\exp(im\Omega_p t)$.  The time-dependent terms in equation
(\ref{eq:LBK1}) then take the simple form
\begin{eqnarray}
  \left\{\int^t_0 dt^\prime e^{i\ldo (t^\prime - t)}
  a(t^\prime) \right\} a^\ast(t) &=&
  e^{-i\ldo t}
  \left\{\int^t_0 dt^\prime e^{i\ldo t^\prime}
    a(t^\prime) \right\} a^\ast(t)
  \nonumber \\
  &=& a(0)a^\ast(0) 
  { \sin[(\ldo - m\Omega_p)t] \over \ldo - m\Omega_p }
  \nonumber \\
  &\longrightarrow& a(0)a^\ast(0) \pi\delta(\ldo - m\Omega_p)
  \label{eq:lbkconverge}
\end{eqnarray}
where the final expression obtains in the limit $t\rightarrow\infty$.
Upon substituting into equation (\ref{eq:LBK1}), we recover the LBK
formula.  The mathematics of this convergence also indicates a
possible numerical pitfall in applying equation (\ref{eq:dfser4}):
at large times $T$, equation (\ref{eq:lbkconverge}) will oscillate
rapidly with $E$ and $J$ or $\bI$ and may require special treatment.

Equation (\ref{eq:LBK1}) can be integrated to find the run of angular
momentum $L_z$ with time for some arbitrary time-dependent
perturbation.  At each time step, one must expand $H_1(\bI, t)$ for
the desired range of $\bl$, integrate over the temporal history, and
then integrate over phase space.  The solution for a general time
dependence is computationally intensive, therefore. In the slowing-bar
example ($\S\ref{sec:barslow}$), the problem has separable time
dependence of the form $H_1(\bI,\bw, t) = a(t) H_1(\bI, \bw)$.  In
this equation (\ref{eq:LBK1}) may be written
\begin{eqnarray}
  \left\langle {dL_z\over dt}\right\rangle &=& -(2\pi)^3
  \sum_{l_1,l_2,l_3=-\infty}^{\infty} l_3 \int\int\int {dE dJ J
  d(\cos\beta) \over \Omega_1(E, J)}\ \bl\cdot{\partial
  f_o\over\partial\bI} \times \nonumber \\
  && \left\{\int^t_0 dt^\prime e^{i\ldo
  (t^\prime-t)} a(t^\prime) \right\} a^\ast(t) H_{1\,\bl}(\bI)
  H_{1\,-\bl}^\ast(\bI).
  \label{eq:LBK2}
\end{eqnarray}
In this case, a grid of $H_{1\,\bl}(\bI)$ along with the other
phase-space dependent terms may be determined once and tabled for all
time, simplifying the calculation.  The integral $\int^t_0 dt^\prime
\exp[i\ldo (t^\prime-t)] H_{1\,\bl}(\bI, t^\prime)$ must be computed
with sufficient small time steps to resolve the oscillations obtained
with variations in $\bI$.  Finally, the right- and left-hand sides of
equation (\ref{eq:dfser4}) are generally coupled.  For a bar--halo
interaction with a fixed bar shape for example, the pattern speed is
$\Omega_p=L_z/I_z$ where $I_z$ is the bar's moment of inertia.
Stability of the solution is then a concern.  I used Gear's method
\cite{Gear:69} to maintain stability for the numerical solutions
presented here. The problem is numerically more stable when the right
hand side is independent of $L_z$ as is the case for a satellite--halo
interaction whose orbital decay is computed from local Chandrasekhar
dynamical friction ($\S\ref{sec:sinksat}$).

\section{Examples}
\label{sec:examples}

\subsection{Slowing bar}
\label{sec:barslow}

Following Paper 2, we compare the results of N-body simulations and
linear theory for an ellipsoidal bar in an NFW dark-matter halo
profile \citep{NFW:97}: $\rho(r)\propto r^{-1}(r + r_s)^{-2}$.  The
fiducial model has a bar of length (semi-major axis) $r_s$ with and
axis ratios $b/a=0.2, c/a=0.05$ and bar mass equal to one-half of the
enclosed dark-halo mass.  The total mass of the bar is 0.1 of the halo
mass inside the virial radius for a concentration $c=15$.  A realistic
present-day stellar bar would have a length roughly 25\% of $r_s$.
However, this length greatly increases the number of particles
required in the N-body simulation.  Because the NFW profile is scale
free inside of $r_s$, the behaviour should scale to smaller bars.  This
is verified in Paper 2.

The pattern speed evolves while conserving the total angular momentum
of the initial bar and dark halo system.  The bar perturbation is
slowly turned on according to $f(t)= (1+\erf[(t-T_0)/\Delta])/2$ with
$T_0=1/2$ Gyr and $\Delta=1/4$, which corresponds to a full-width
turn-on time of 1 Gyr scaled to the Milky Way.  Bar formation in
simulations occurs over a small number of dynamical timescales
(e.g. within 200 Myr scaled to the Milky Way).  An assumption of
$T_0=1/2$ and $\Delta=1/4$ then is an upper bound to a realistic
formation time scale; we are choosing the slowest time dependence
possible, accounting for bar length scale.  Paper 2 describes the
additional details of the simulation procedure.

\begin{figure}
  \centering
  \subfigure[NFW]{\includegraphics[width=0.8\linewidth]{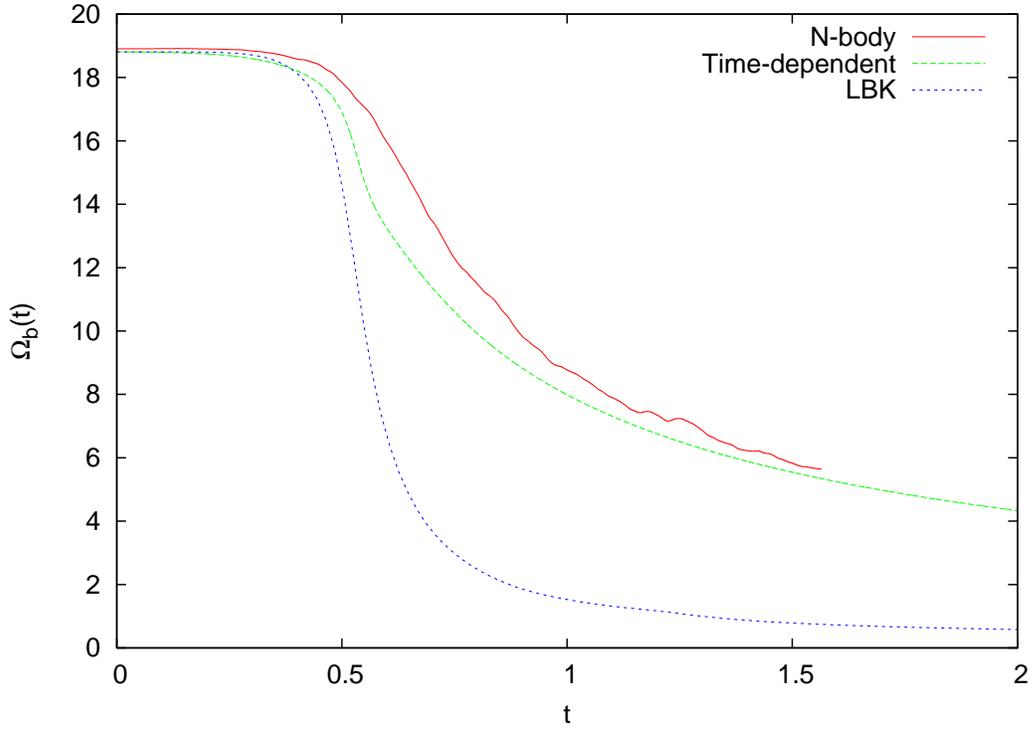}}
  \subfigure[King]{\includegraphics[width=0.8\linewidth]{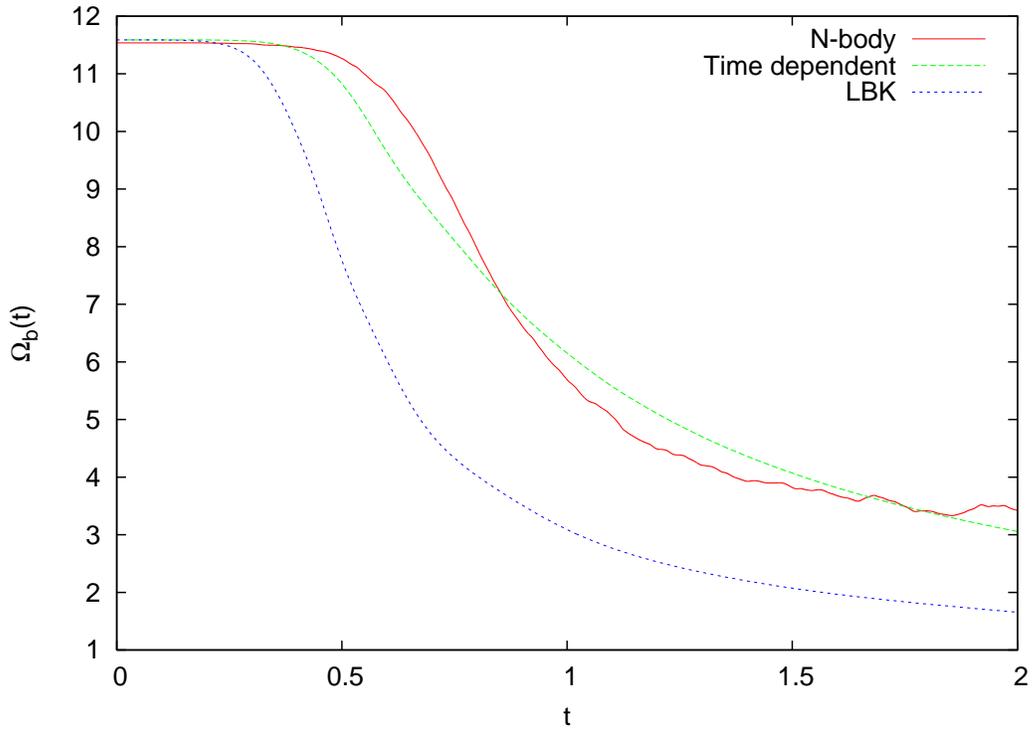}}
  \caption{The pattern speed evolution of a bar in a
    NFW (a) and King model (b) dark-matter halo for an N-body
    simulation and computed from the second-order perturbation theory
    in the time-dependent and LBK limits.  The time-dependent theory
    accurately follows the patter speed evolution while the
    time-asymptotic LBK theory does not.}
\label{fig:pattern}
\end{figure}

Figure \ref{fig:pattern} compares the resulting evolution of the
N-body simulation, the linear theory presented in \S\ref{sec:method}
and the LBK formula.  Because the bar figure is constant in time, the
torque takes the form of equation (\ref{eq:LBK2}).  The N-body
simulation closely follows the time-dependent linear theory
prediction.  The evolution is significantly slower than predicted by
the LBK formula!  Clearly, the time-dependence changes the torque.
The second panel of Figure \ref{fig:pattern} repeats the simulation
and linear theory prediction for a bar in a King model profile
\citep{King:66} chosen to match the cumulative mass profile of the NFW
profile inside of the virial radius ($W_0=6$, $r_T=2 r_{vir}$)
demonstrating that the differences due to time dependence are not a
consequence of the NFW profile.

\begin{figure}
  \centering
  \subfigure[LBK]{\includegraphics[width=0.8\linewidth]{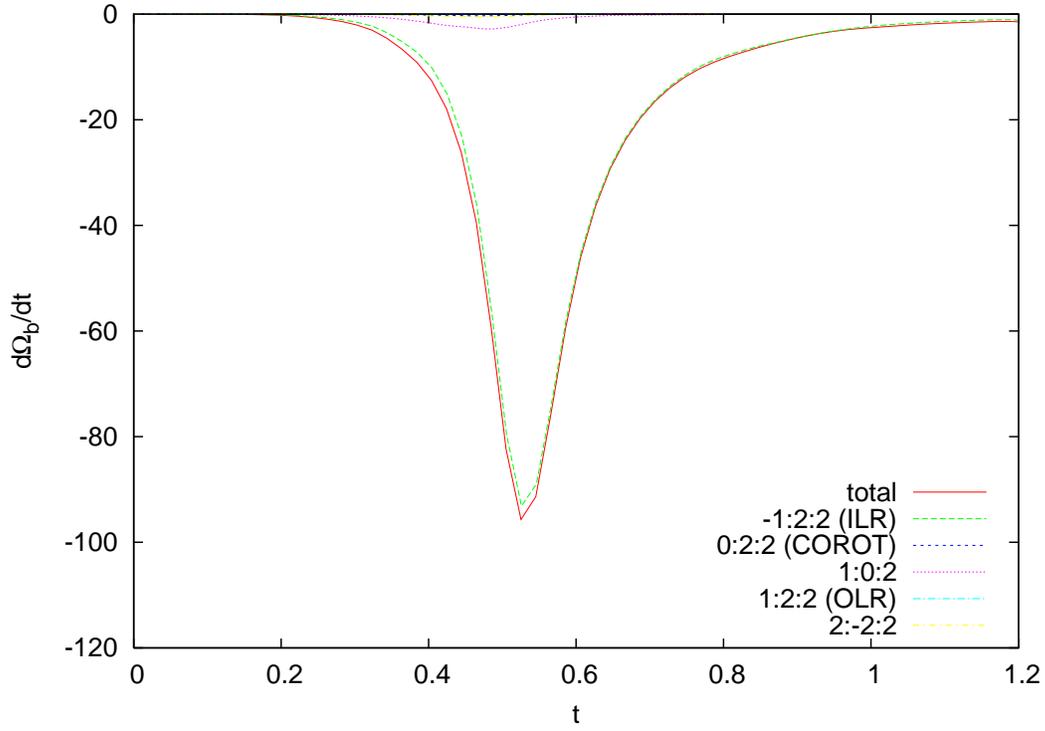}}
  \subfigure[time-dependent]{\includegraphics[width=0.8\linewidth]{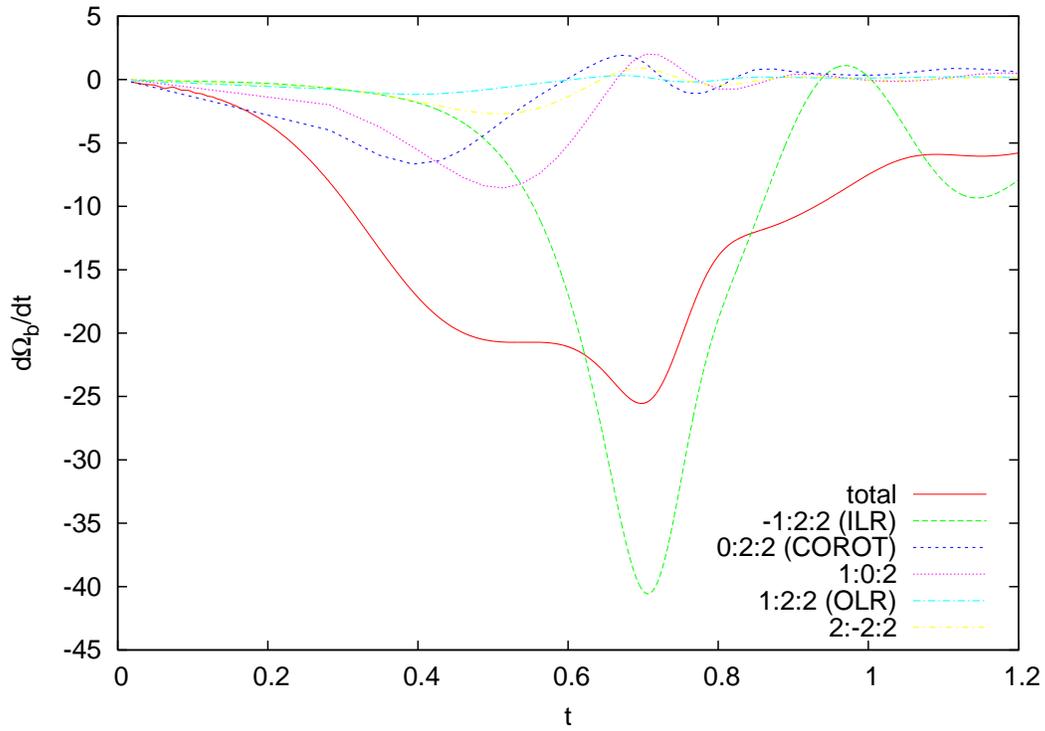}}
  \caption{Relative contributions to the
    total torque ${\dot L}_z = I_z{\dot\Omega}_z$ from each resonance
    for time-asymptotic (LBK) regime (a) and with explicit
    time-dependence (b).  Each curve in the figure shows a single
    $\bl$ term from equation (\protect{\ref{eq:LBK1}}) as labelled.
    The run of ${\dot\Omega}_z$ derived from the total torque is also
    shown.}
\label{fig:lbk}
\end{figure}

Equation (\ref{eq:LBK1}) or (\ref{eq:LBK2}) allows us to determine the
contributions from each resonance separately (see Fig. \ref{fig:lbk})
and determine the origin of the differences between the time-dependent
theory and LBK.  The qualitative differences between the case with
explicit time-dependence and the time-asymptotic LBK regime are
evident.  For the LBK case, the torque is dominated by the ILR-like
$\bl=(-1,2,2)$ resonance.  The contribution peaks as the bar begins to
slow and decreases when the pattern frequency becomes too small to
easily couple to low-order resonances in the vicinity of the bar.  For
the time-dependent case, the corotation resonance $(0,2,2)$ dominates
at early times.  During the rapid slowing phase, $0.4\la t\la 1$, the
ILR $(-1,2,2)$ dominates.  The ILR is weaker in the time-dependent
theory and the simulation.  This accounts for a large fraction of the
difference in Figure \ref{fig:pattern}.  At later times, a near
equilibrium obtains between the various commensurabilities, some
without true resonances.  These all converge to the LBK limit as
$t\rightarrow\infty$.

\begin{figure}
\centering
\includegraphics[width=0.9\linewidth]{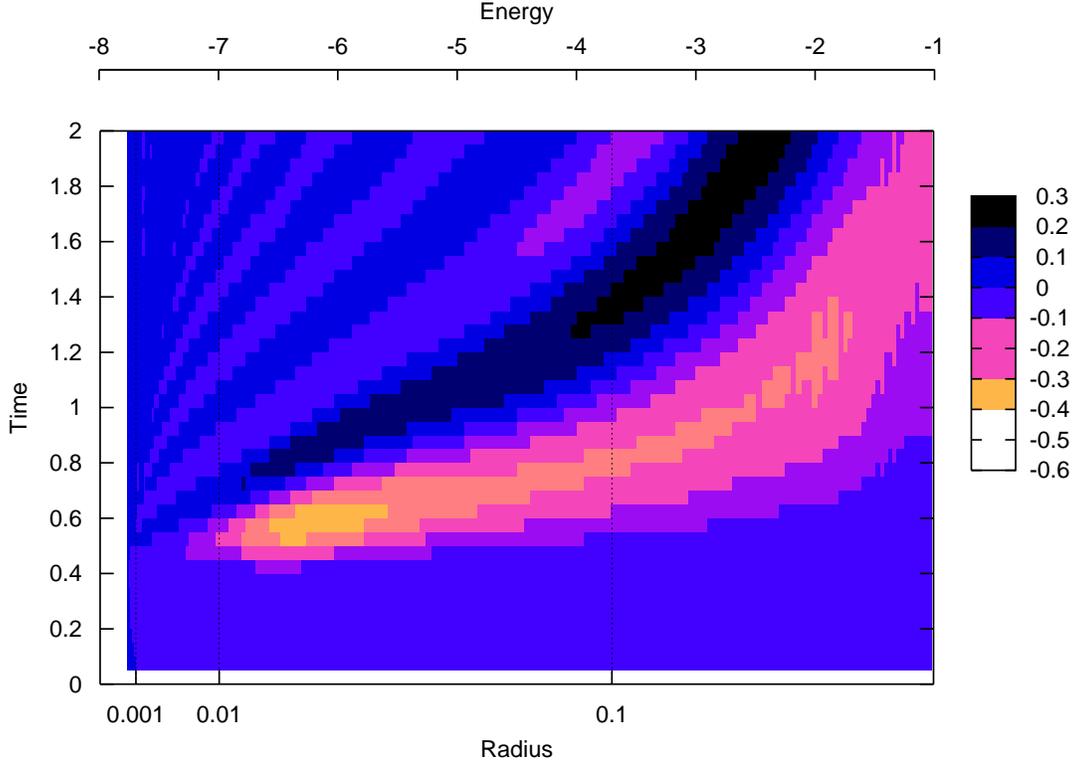}
\caption{ILR contribution to the torque (${\dot\Omega}_z={\dot
  L}_z/I_z$) as a function of energy (top axis) and corresponding
  radius of the corresponding circular orbit (bottom axis) as a
  function of time as the pattern speed evolves.  This figure is
  obtained by solving equation (\protect{\ref{eq:LBK1}}) and plotting
  the energy contribution to the phase-space integral as a function of
  time for the same example shown in Fig. \protect{\ref{fig:lbk}}.}
\label{fig:energyevolve}
\end{figure}

In Figure \ref{fig:energyevolve}, I show the torque from the ILR
summed over all angular momentum plotted as a function of energy $E$
(top axis) and radius $r$ of the circular orbit with the same energy
$E$ and time $t$ (bottom axis).  The peak angular momentum deposition
occurs inside the bar radius (due to the ILR) and corresponds with the
radial location of the profile change described in Paper 2.
Comparison of Figures \ref{fig:lbk} and \ref{fig:energyevolve}
confirms that the peak torque occurs during the rapid pattern speed
evolution which also accounts for the broad region of this
contribution in energy for $0.4\la t\la 1$.  For $t\ga 1$, the ILR
oscillates at lower amplitude as it approaches the constant
$t\rightarrow\infty$ LBK value.

\begin{figure}
\centering
\includegraphics[width=0.9\linewidth]{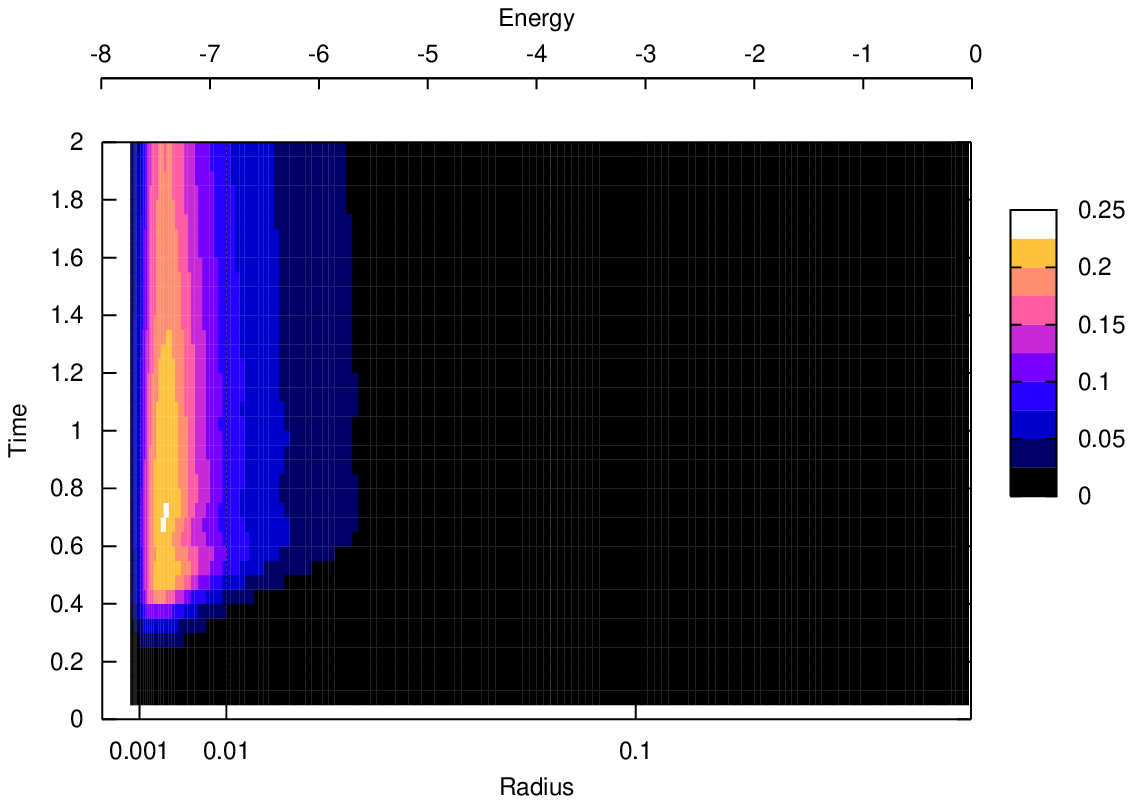}
\caption{Ratio of the cumulative angular momentum contributed by the
  ILR inside of energy $E$ and radius $r$ of the corresponding
  circular orbit at time $t$ to the total angular momentum of the halo
  inside of energy $E$.  In other words, the contour scale shows the
  fractional change of the angular momentum in the halo from the ILR
  resonance inside of a given energy or radius.}
\label{fig:cumtorque_bar}
\end{figure}

\begin{figure}
\centering
\subfigure[$a=0.067$, $M_b=0.01$, $T_o=1/2$, $\Delta=1/4$]{%
  \includegraphics[width=0.45\linewidth]{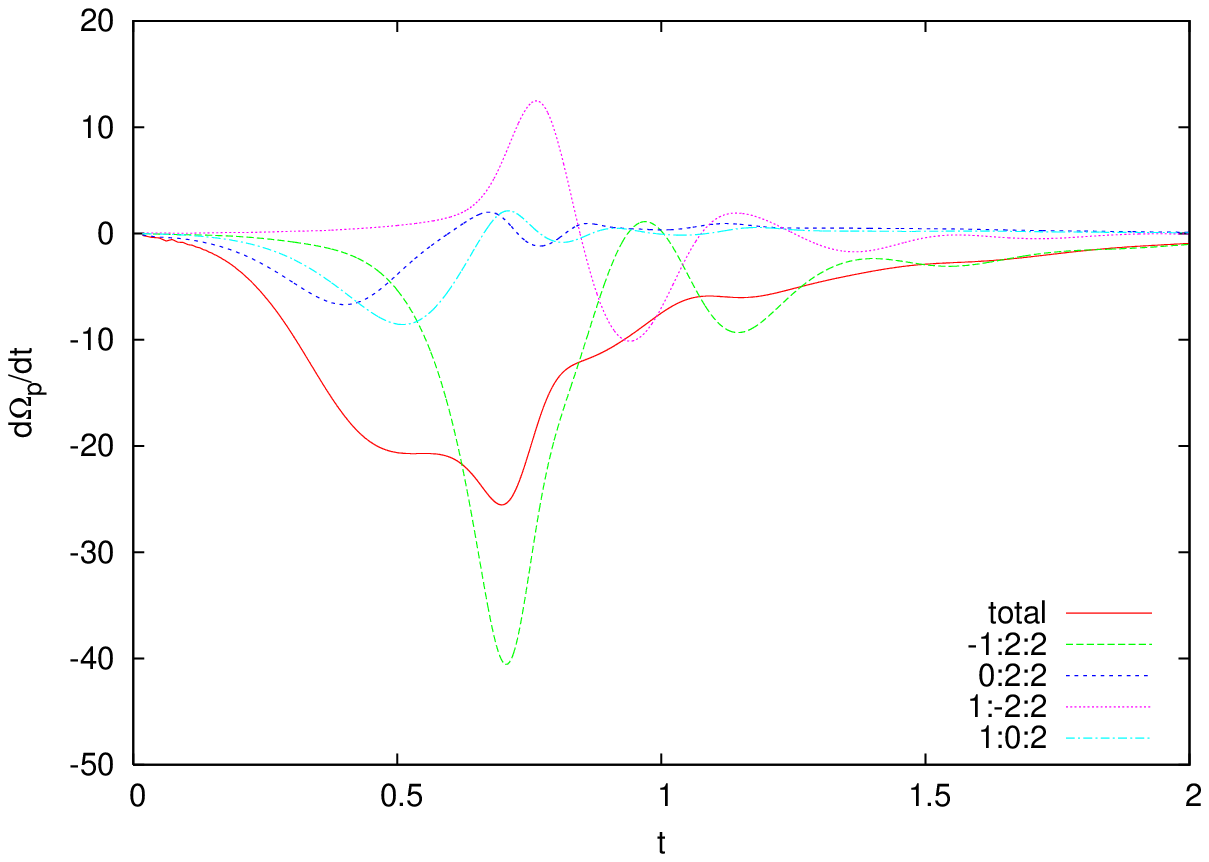}
}
\subfigure[$a=0.067$, $M_b=0.01$, $T_o=1$, $\Delta=1/3$]{%
  \includegraphics[width=0.45\linewidth]{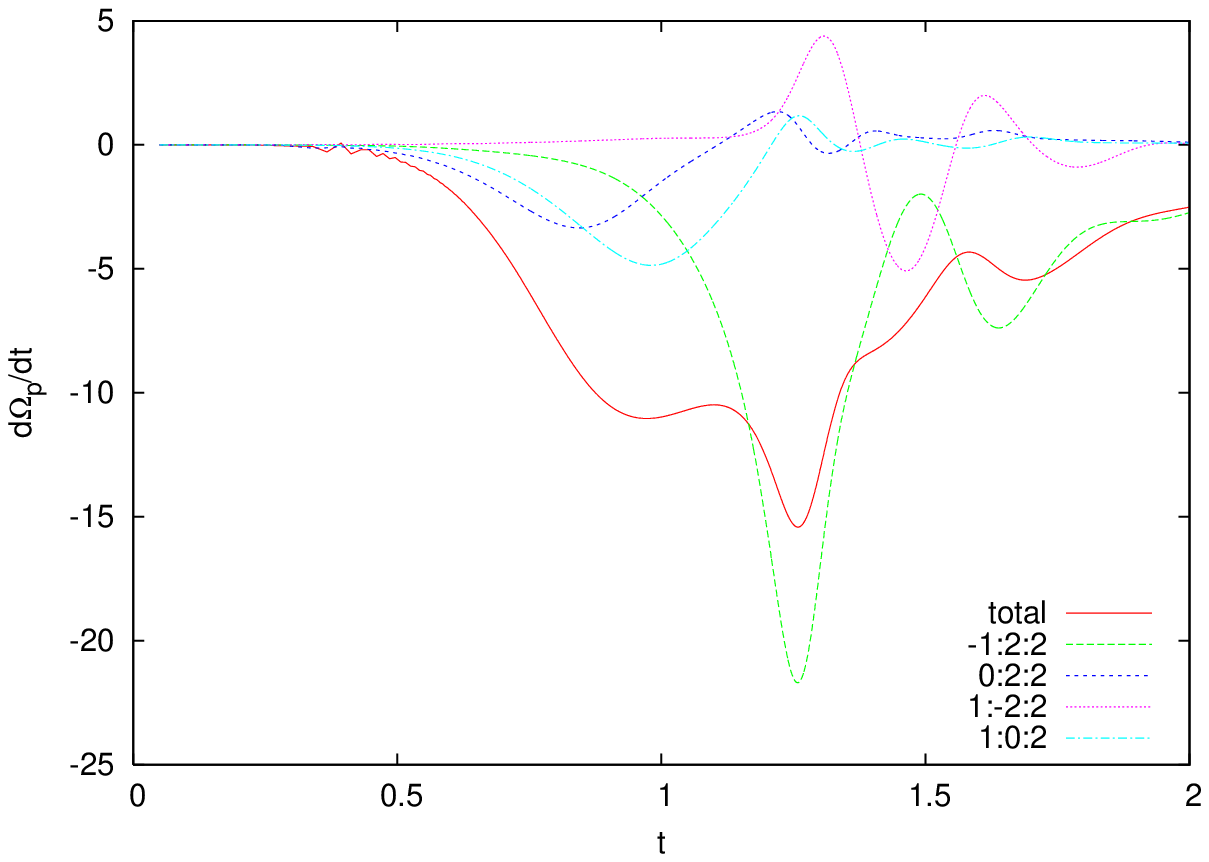}
}
\subfigure[$a=0.067$, $M_b=0.01$, $T_o=3/2$, $\Delta=1/2$]{%
  \includegraphics[width=0.45\linewidth]{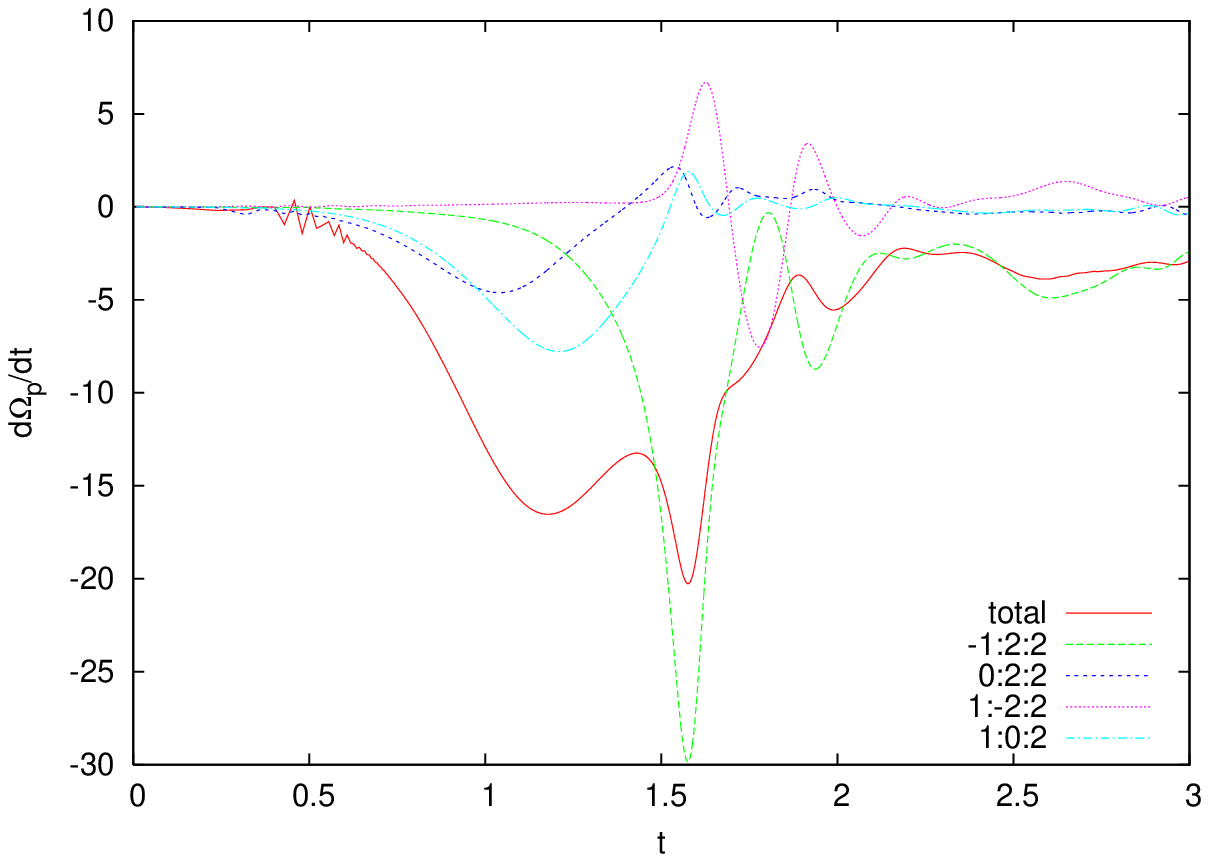}
}
\subfigure[$a=0.02$, $M_b=0.015$, $T_o=1/2$, $\Delta=1/4$]{%
  \includegraphics[width=0.45\linewidth]{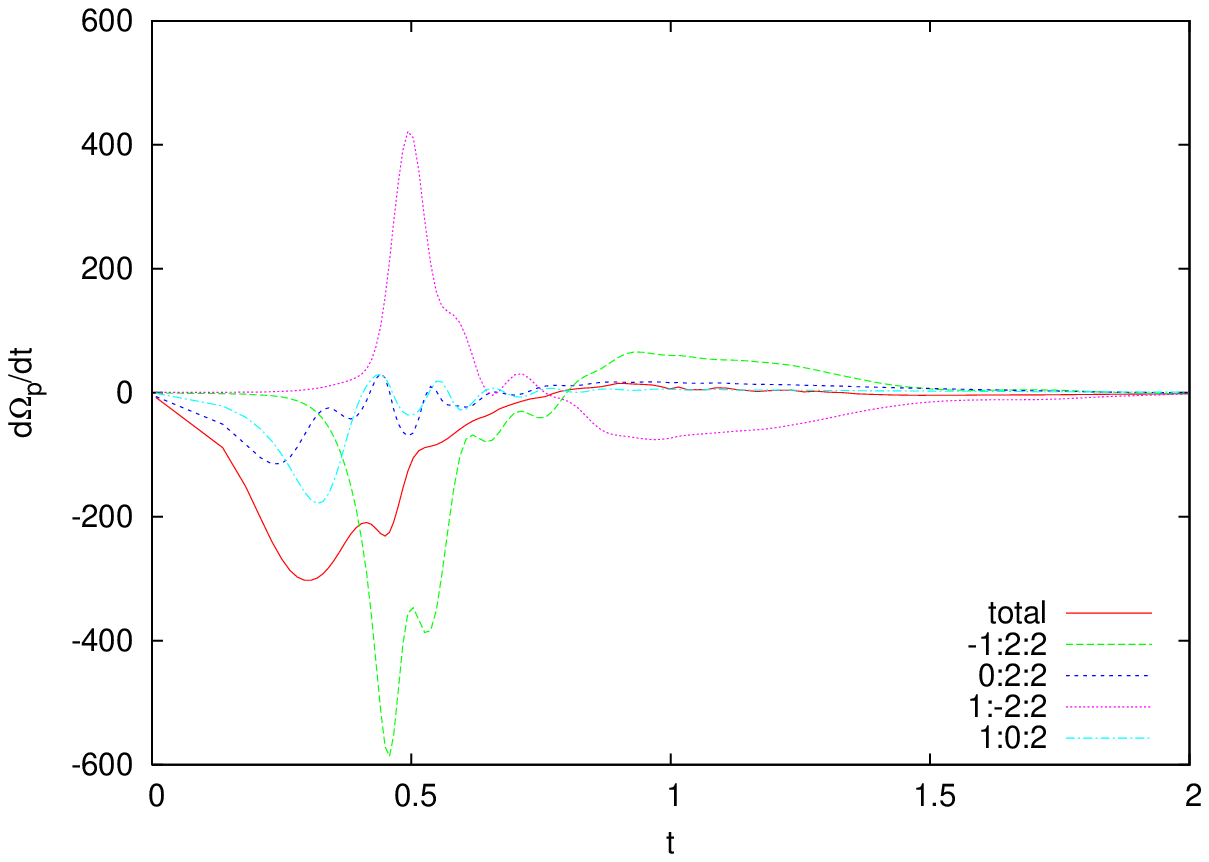}
}
\subfigure[$a=0.01$, $M_b=0.011$, $T_o=1/2$, $\Delta=1/4$]{%
  \includegraphics[width=0.45\linewidth]{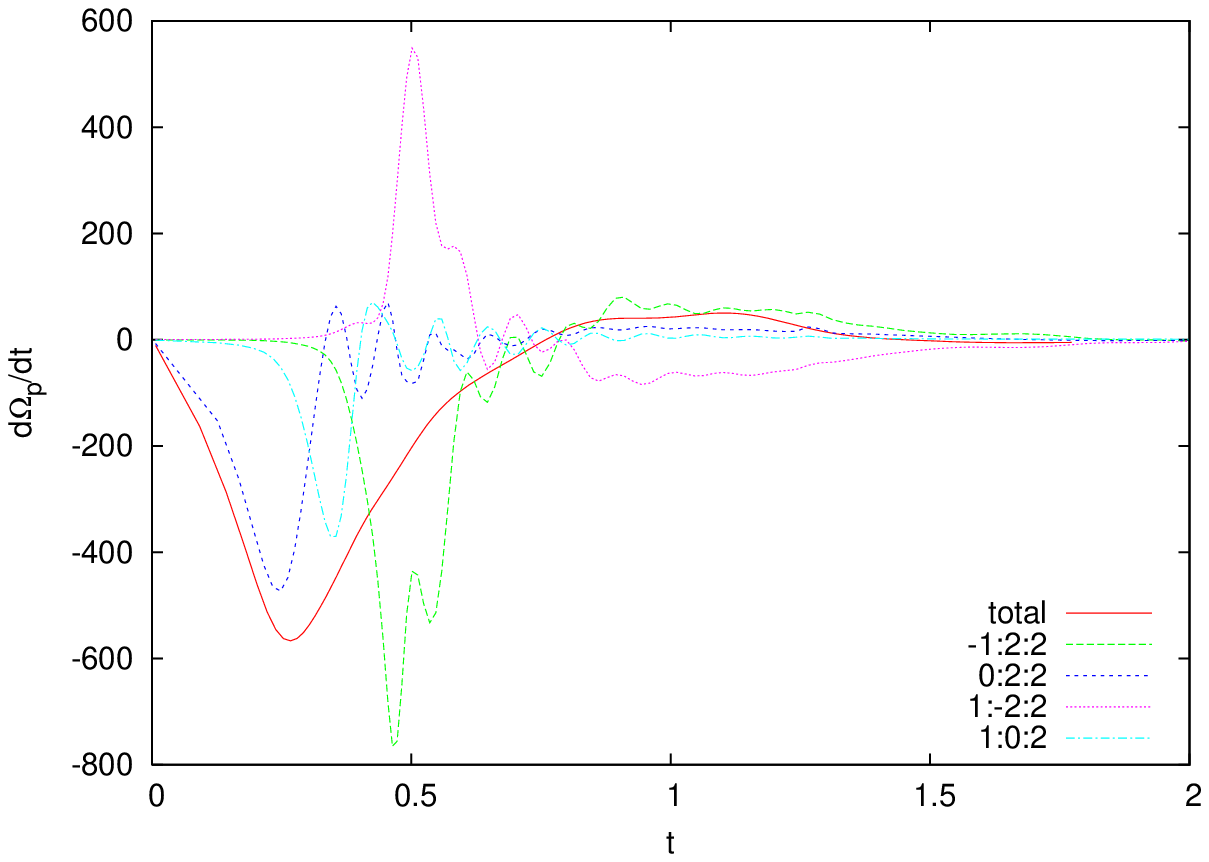}
}
\caption{Relative contributions to the
  total torque ${\dot L}_z = I_z{\dot\Omega}_z$ as in Fig.
  \protect{\ref{fig:lbk}} for a range of bar lengths ($a$), bar masses
  ($M_B$) and growth times ($T_0$ and $\Delta$).  Only the top four
  contributors are shown individually.  The total torque is the sum
  over all contributing terms for $l=m=2$.}
\label{fig:restorque}
\end{figure}

We now take a closer look at the contributions from all terms $(l_1,
l_2, l_3=m)$ in equation (\ref{eq:LBK2}).  The contribution to the
torque for each term and the total is shown in Figure
\ref{fig:restorque} for a variety of different bar lengths and turn-on
parameters $T_0, \Delta$.  The mass of the bar has been decreased by a
factor of five to $M_b=0.01$ to lengthen the slow down period and help
isolate the contribution of each resonance. For clarity, only the top
four contributing resonances are drawn.  The first panel of the
describes the torque for a bar with length $r_s$ with initial period
$1/3$, $T_0=1/2$ and $\Delta=1/4$; the bar grows to full strength
over several bar periods.  At $t<1/2$, early in the bar growth, the
corotation $(0,2,2)$ and the $(1,0,2)$ resonances dominate the torque
(see also Fig.  \ref{fig:lbk}). At the peak of the pattern-speed
decrease, $t\approx0.7$, the ILR dominates the torque. Moreover, the
ILR dominates to the integrated torque for the entire slowdown event.
Note that terms $\bl$ without pattern-speed commensurabilities,
$l_1\Omega_1+l_2\Omega_2\not=m\Omega_p$, through the phase-space
distribution also contribute transiently but significantly to the
total torque.  For example, the $(1,-2,2)$ term, an ``negative'' ILR
contributes to the torque but would only be commensurate for a
negative pattern speed.  This transient echo decays after the rapid
slow down phase since it is purely oscillating without a net
second-order contribution.  Nonetheless, it diminishes the torque
during peak of the slow down.  The ILR and the other commensurate
terms, on the other hand, approaches the finite non-zero
time-asymptotic value predicted by the LBK formula.

The relative importance of the torque to the inner halo is nicely
visualised by computing the net change in angular momentum inside of
some energy $E$ from equation (\ref{eq:LBK1}) as a function of time.
That is, let ${\dot L}(<E)$ denote equation (\ref{eq:LBK1}) with the
energy integral from $[E_{min}, E]$.  Then, define the cumulative
change of angular momentum in time as
\begin{equation}
\Delta L(<E, <t) = \int_{t_o}^t dt {\dot L}(<E)
\end{equation}
where $t_o$ is the initial time.  The total orbital angular momentum
in the initial phase space with energy less than $E$ is
\begin{equation}
  {\cal E}(<E) = (2\pi)^3 \int\int\int {dE dJ J
    d(\cos\beta) \over \Omega_1(E, J)} f(E, J) J.
\end{equation}
Then, the ratio of these two quantities, $\Delta L(<E, <t)/{\cal
E}(<E)$ indicates the importance of the resonant torque to the
structure of the inner halo profile.  This ratio is shown in Figure
\ref{fig:cumtorque_bar} for the bar--halo interaction depicted in
Figure \ref{fig:energyevolve} for the ILR resonance.  The contour
scale shows the fractional change of the angular momentum in the halo
inside of a given energy or radius.  A large fractional change
indicates a strong possibility of significant structural change.  The
peak {\em absolute} torque from the ILR occurs near $r\approx0.02$ and
$t\approx0.6$ (Fig. \ref{fig:energyevolve}) while the peak {\em
relative} torque occurs near $r\approx0.003$
(Fig. \ref{fig:cumtorque_bar}).  In particular, N-body simulations
from Paper 2 show that this bar--halo interaction will flatten the NFW
cusp for $0.001<r<0.01$ as predicted by Figure
\ref{fig:cumtorque_bar}.  Hence, this comparison suggests that a
15\%-20\% fractional change is sufficient to drive inner-halo
evolution.

The sequence in Panels (a)--(c) in Figure \ref{fig:restorque} shows
the same bar model with different turn on parameters $T_0$ and
$\Delta$.  The corresponding distribution of torque deposited as a
function of time and energy (radius) is similar to Figure
\ref{fig:energyevolve}.  For longer intervals of bar growth, the
resonances are more localised in energy and radius.  The steep ILR
peak during the rapid evolution phase covers a broad swath in energy
and radius and leads to profile evolution.  At later times as the
system approaches the time-asymptotic phase, the torque contributions
become smaller and more localised.  Transients remain important for
the unrealistically slow turn on $\Delta=1/2$ (full width of 2 Gyr
scaled to the Milky Way).  The profiles for smaller, more realistic
size bars are qualitatively similar are similar to those for the large
bar (Panels d \& e in Fig.  \ref{fig:restorque}); ILR continues to
dominate the total torque with corotation and $(1,0,2)$ important at
early times.  For these smaller bars, the ILR changes sign after the
main peak ($T\approx1$) as the initial response laps the slowing bar
before settling to the time-asymptotic LBK value.  The torque is
deposited as smaller radii and energy as expected but still over a
broad region inside of the bar radius.  Paper 2 presents estimates of
the particle number required to simulate the resonant interaction in
these examples.

\subsection{Decaying satellite}
\label{sec:sinksat}

\begin{figure}
  \centering
  \includegraphics[width=0.75\linewidth]{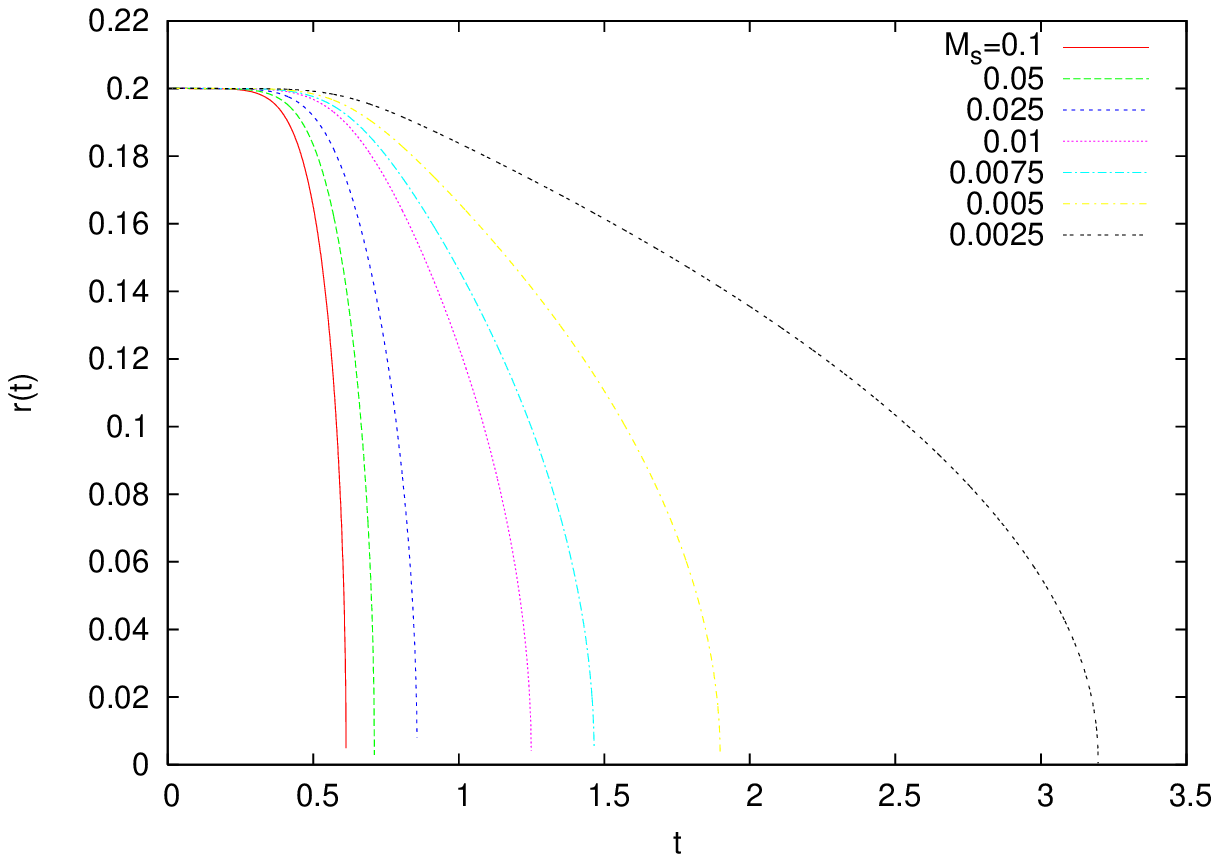}
  \caption{Change in radii with time due to orbital decay according to
    the local Chandrasekhar dynamical friction formula for nearly
    circular orbits for a variety of satellite masses $M_s$ (in units
    of the halo virial mass.}
  \label{fig:lbksatorb}
\end{figure}

\begin{figure}
\centering
\subfigure[$M_s=0.025$]{%
  \includegraphics[width=0.45\linewidth]{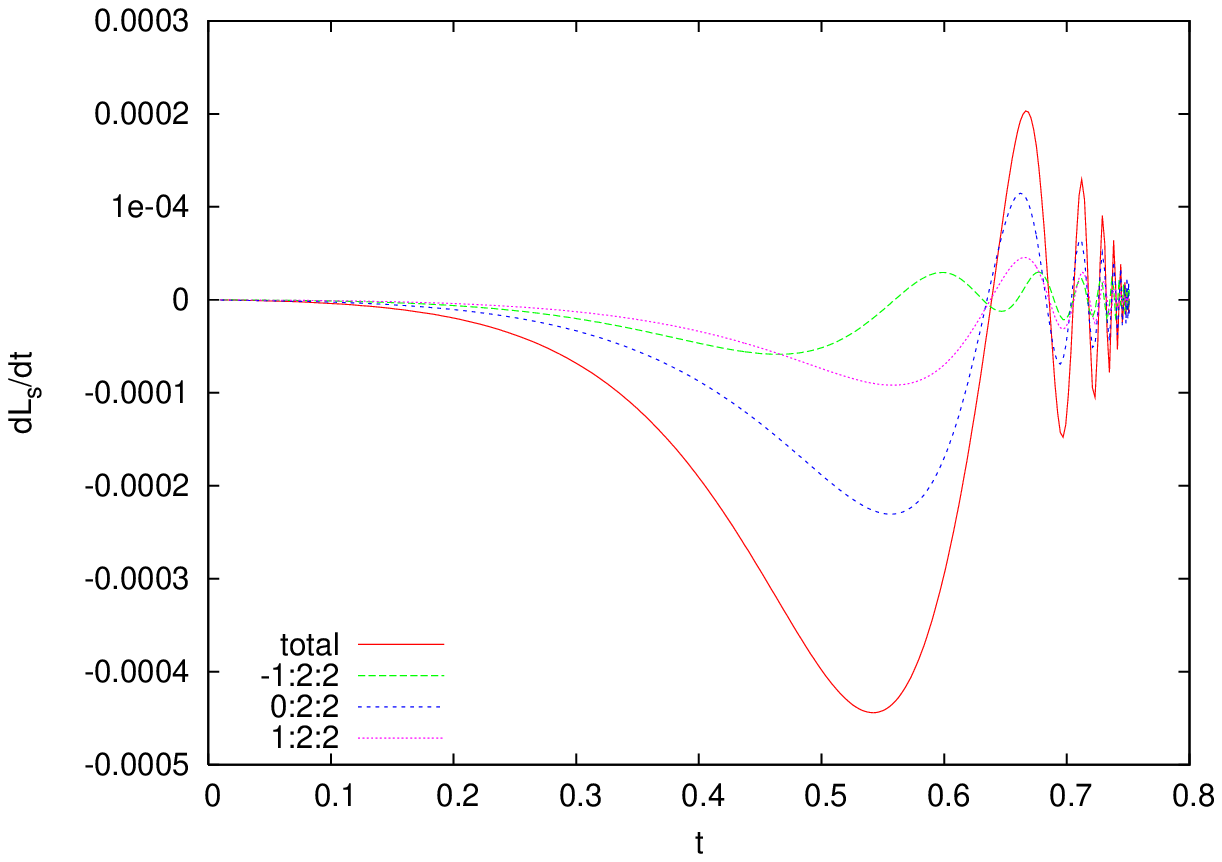}
}
\subfigure[$M_s=0.025$]{%
  \includegraphics[width=0.45\linewidth]{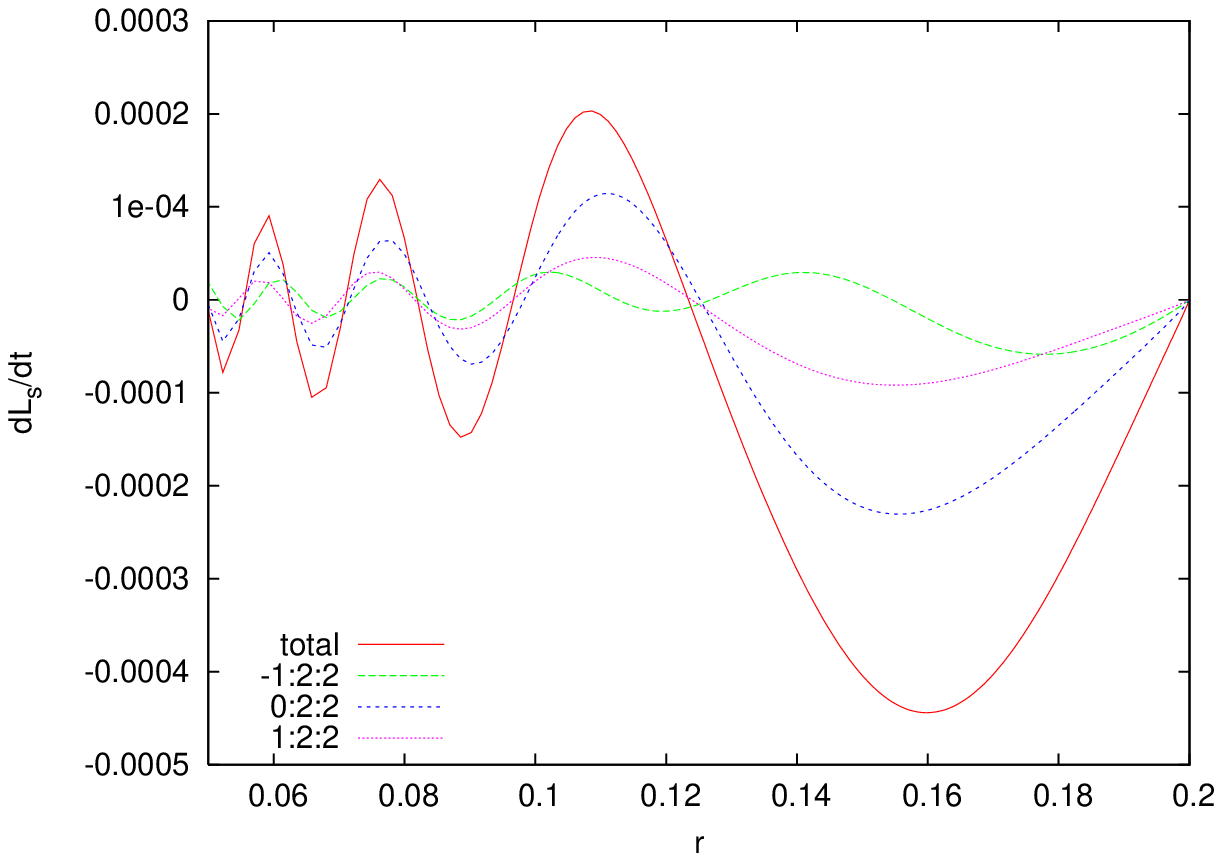}
}
\subfigure[$M_s=0.01$]{%
  \includegraphics[width=0.45\linewidth]{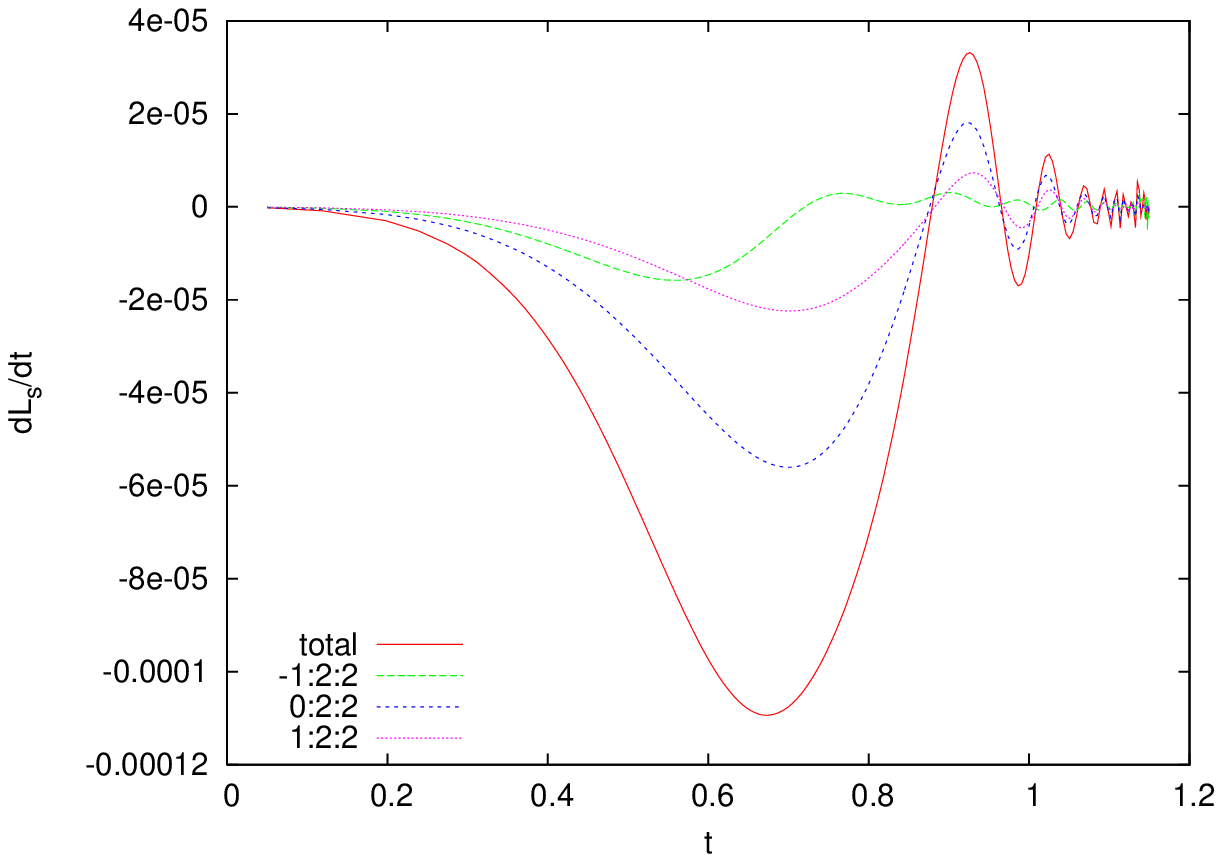}
}
\subfigure[$M_s=0.01$]{%
  \includegraphics[width=0.45\linewidth]{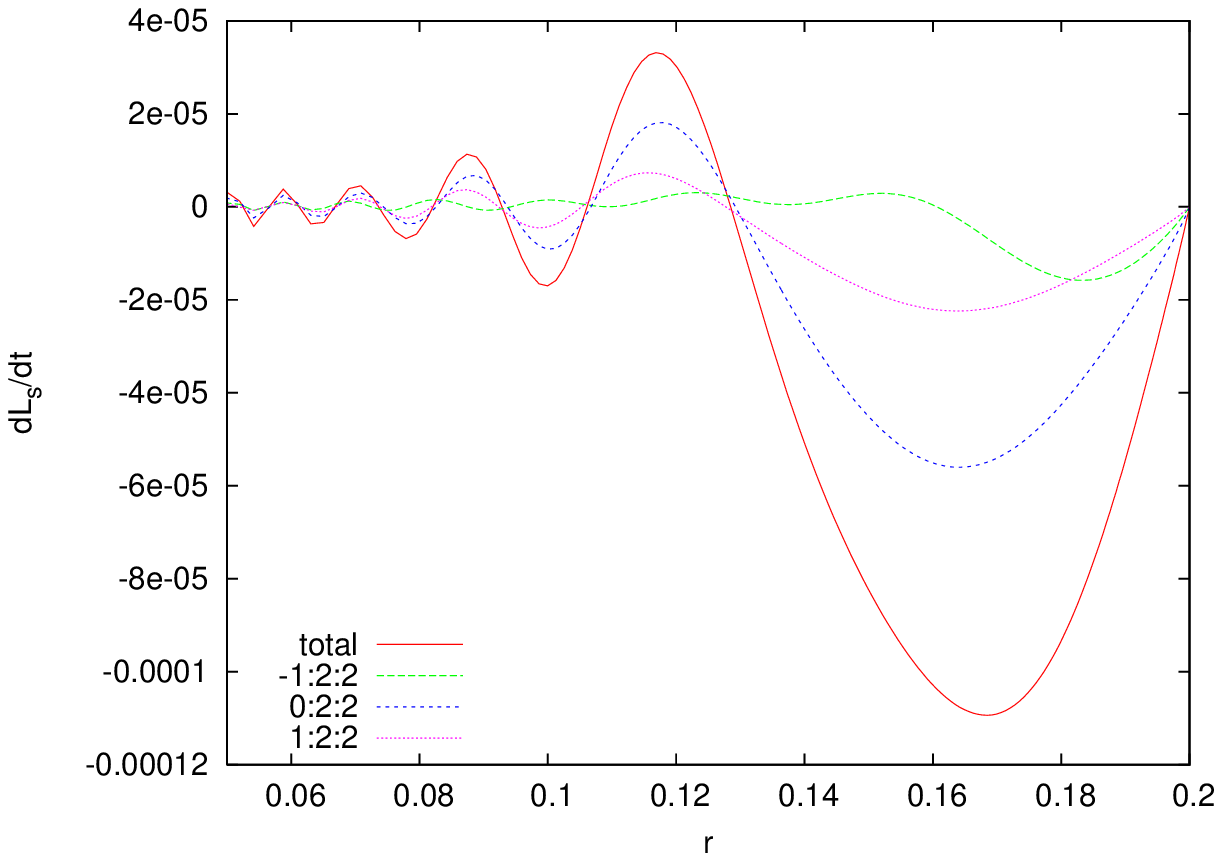}
}
\subfigure[$M_s=0.005$]{%
  \includegraphics[width=0.45\linewidth]{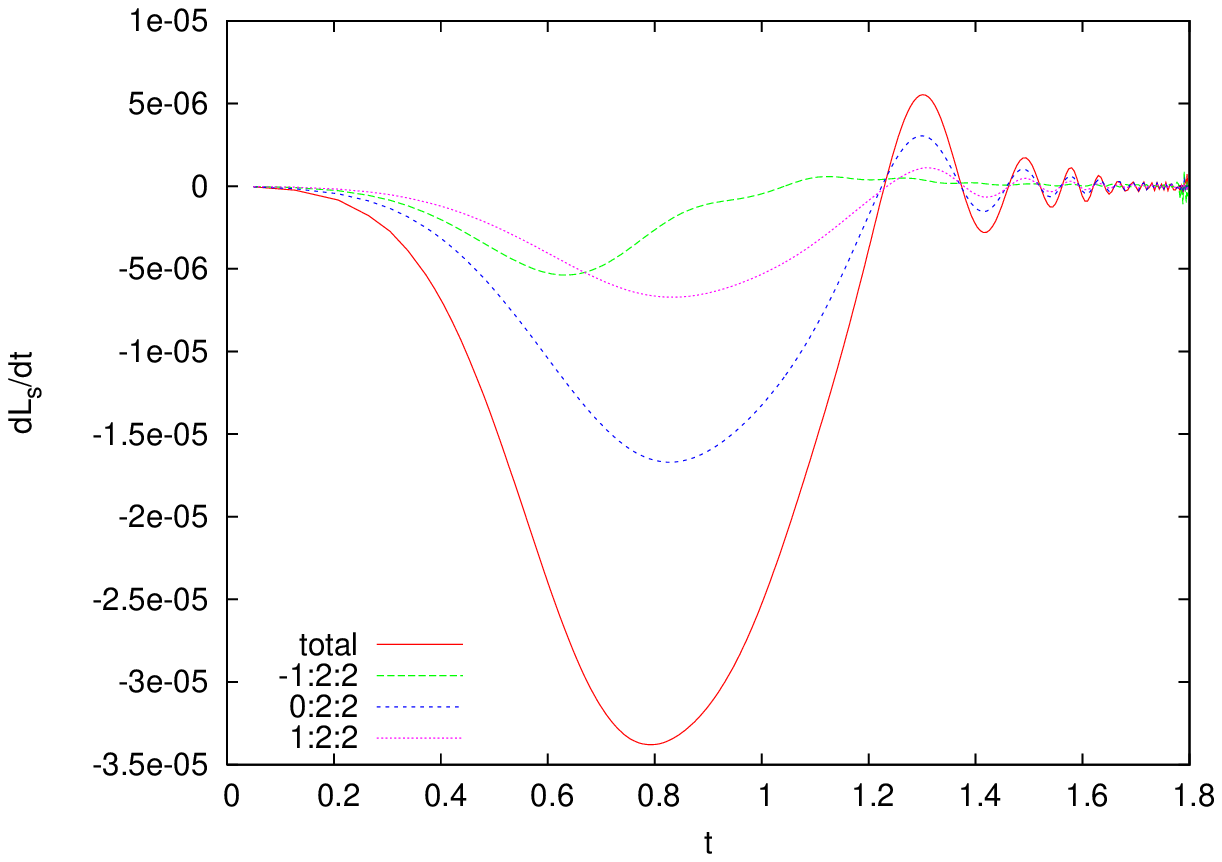}
}
\subfigure[$M_s=0.005$]{%
  \includegraphics[width=0.45\linewidth]{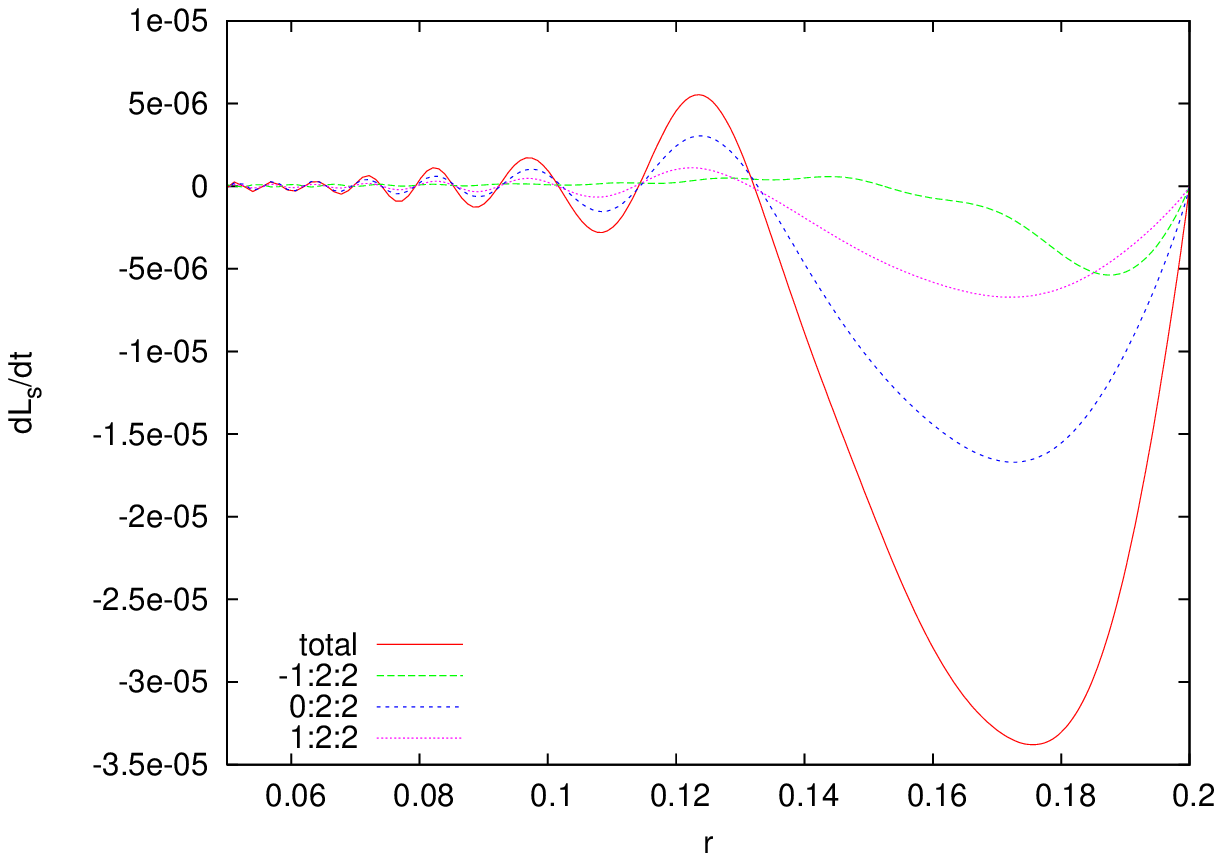}
}
\caption{Time-dependent torque for various resonances for a decaying
  circular satellite orbit with $\log\Lambda=4$ $T_o=1/2$,
  $\Delta=1/4$, and initial radius $R_0=0.2$.  Only the top three
  contributing terms are shown individually.}
\label{fig:lbksat}
\end{figure}

\begin{figure}
  \centering
  \includegraphics[width=0.9\linewidth]{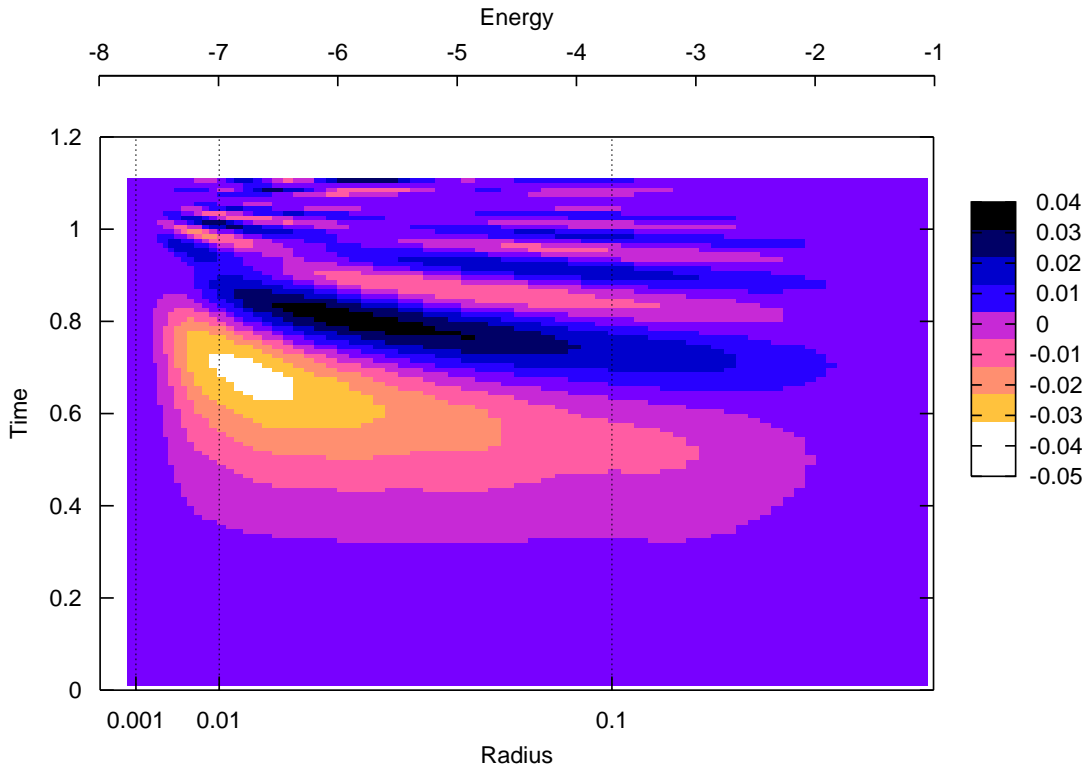}
  \caption{ILR contribution to the torque as in
    Fig. \protect{\ref{fig:energyevolve}} but for the satellite decay
    with $M_s=0.025$.}
  \label{fig:ilrlzsat}
\end{figure}

\begin{figure}
\centering
\includegraphics[width=0.9\linewidth]{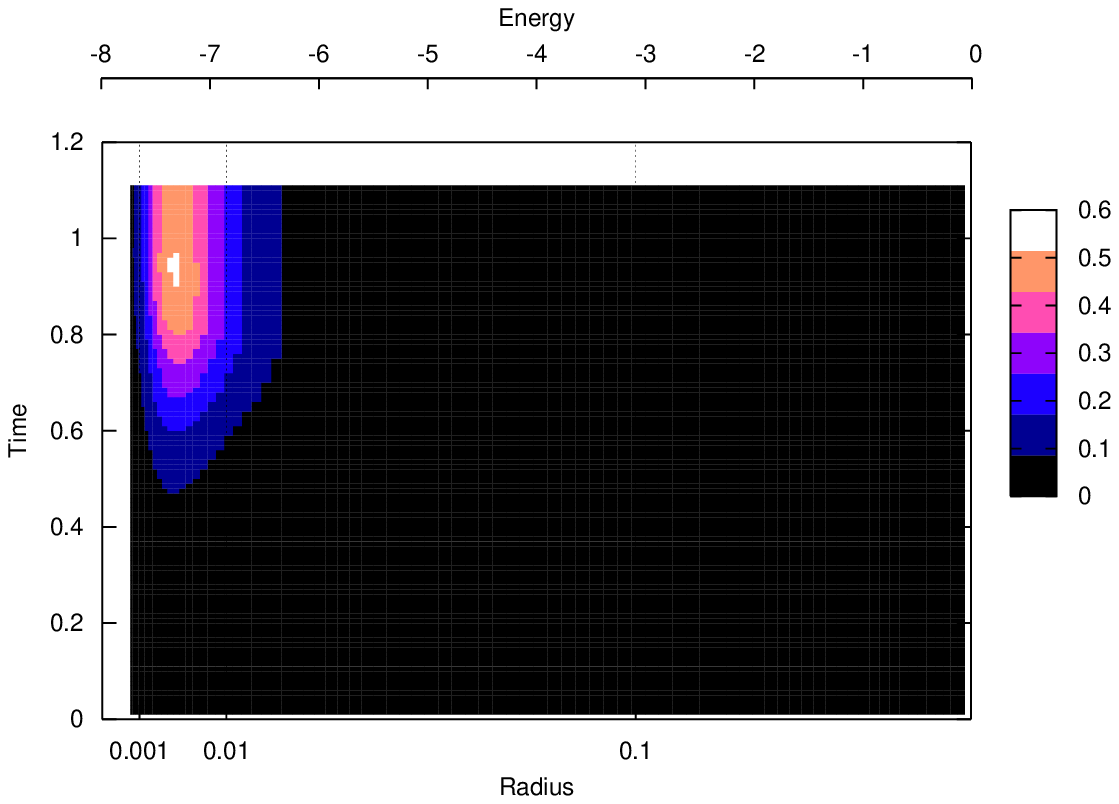}
\caption{Ratio of cumulative ILR angular momentum transfer to total
  angular momentum as in Fig. \protect{\ref{fig:cumtorque_bar}} but for
  the satellite decay with $M_s=0.025$.}
\label{fig:cumtorque_sat}
\end{figure}

\subsubsection{Circular orbits}

Orbiting satellites couple to a {\em live halo} through resonances in
much the same way as does a rotating bar.  However, because the
satellite size is generally much smaller than the halo scale, the
total torque will have significant contributions from harmonic
orders\footnote{The contribution scales as $(\ln l)^{-1}$ where $l$ is
the harmonic order \protect{\citep{Weinberg:86}}.} larger than
quadrupole. Just as in the case of the bar, the $(-1,2,2)$ resonance
can couple to very small radii although the satellite perturbation
will have a different frequency spectrum and amplitude than the bar
perturbation. Higher-order contributions have energies and response
locations similar to the satellite.  To investigate this situation, I
assume that the satellite's orbit decays according the the local
Chandrasekhar formula while remaining on a circular orbit.  This, then
determines the radius of the satellite orbit and the instantaneous
orbital frequency, $\Omega_s$.  The low-order response is then
evaluated using the theory from \S\ref{sec:method}.  It is possible to
compute the torque self-consistently by including a larger number of
$l$ terms (eq. \ref{eq:dfser4}).  However, in this example, we only
want to illustrate the contribution of the lower-order terms for
comparison with the bar example and therefore the additional
complexity and time to compute the self-consistent evolution is not
justified.  Because the satellite orbital radius changes in time, the
torque takes the form of equation (\ref{eq:LBK1}): the time-dependence
in the rotation is not separable from the time-dependence of the
changing Fourier coefficients in equation (\ref{eq:ftcoef}).  The
torque is then straightforwardly computed for low-order $l$ and $m$
using equation (\ref{eq:LBK2}) although it is more computational
intensive than the bar example due to the non-separable time
dependence than the bar evolution case.

The initial orbital radius is $r=0.2$ (20\% of the virial radius or
approximately 60 kpc scaled to the Milky Way).  Figure
\ref{fig:lbksatorb} shows the radius as a function of time for a
variety of satellite masses $M_s$ from 0.0025 to 0.1 virial mass units
for interpreting the torque plots that follow.  As in the case of the
bar, the mass is slowly increased using $f(t)$ (from
\S\ref{sec:sinksat}) with $T_0=1/2$ and $\Delta=1/4$.  The smallest
mass satellite requires approximately 6 Gyr to decay from $r=0.2$ and
the most massive satellite decays in less that 100 Myr.

Figure \ref{fig:lbksat} shows the torque, $dL_s/dt$, as a function of
time and orbital radius for the quadrupole $l=m=2$ spherical harmonic
contribution.  The torque for all values of $M_s$ share some common
trends: 1) the $l=m=2$ torque is dominated by corotation (0:2:2).  At
early times, there is a moderate ILR transient (-1:2:2) which changes
sign and oscillates at later times and the satellite position laps the
transient wake, alternatively shifting the sign of the torque.  We can
remove the turn-on and extrinsic amplitude dependence by dividing
$dL_s/dt$ by $M_s^2$ and multiplying the time by
$M_s\int_0^t\,dt^\prime f(t^\prime)$.  Each value of time, then,
corresponds to the same satellite radius in each panel.  The initial
transient in $dL_z/dt$ between radii 0.12 and 0.2 (Panels b, d, and f
of Fig.  \ref{fig:lbksat}) has a nearly self-similar form.  The
differences are due to {\em intrinsic} time dependence from
transients.  For example, as $M_s$ decreases, the ILR becomes more
localised in time with a larger peak amplitude.

Figure \ref{fig:ilrlzsat} illustrates the distribution of torque from
the ILR as a function of time and energy and radius similar to Figure
\ref{fig:energyevolve}.  Torque from the initial transient for the
satellite orbiting at $r\approx0.15\approx 2 r_s$ is deposited at
radii between 0.005 and 0.02 or between 7\% and 30\% of $r_s$.  Figure
\ref{fig:cumtorque_sat} plots the ratio of the cumulative change in
angular momentum in energy and time, $\Delta L(<E, <t)/{\cal E}(<E)$,
for the satellite decay shown in Figure \ref{fig:ilrlzsat}.  As in the
bar--halo interaction, the peak total torque occurs at much larger
radii than the peak relative torque.  Comparing to Figure
\ref{fig:cumtorque_bar}, we see that the change in angular momentum is
in the same region of phase space and larger than the 15\%-20\%
fractional change found in the bar example.  Since the halo evolves
for the bar simulation, we anticipate evolution for the halo driven by
this satellite.  The N-body simulation is more difficult to perform in
this case because of the disparate scales and will be the subject of a
later paper \citep{Choi.etal:04}. Since the decay time is proportional
to $M^{-1}_s$ and the torque is proportional to $M^2_s$, the total
torque deposited in the transient phase is proportional to satellite
mass .  Therefore, as long as $M_s$ is sufficiently large that the
orbital decay takes place in a galaxy age, the angular momentum
transport from a few larger merger events may affect the halo profile
as described in Paper 2.

\subsubsection{Eccentric orbits}

Eccentric orbits may be treated similarly using equation
(\ref{eq:LBK2}).  Rather than enforce a circular orbit, one may solve
the equations of motion for an orbit decaying according to the
Chandrasekhar formula and use the solution, simultaneously, as the
input to equation (\ref{eq:LBK2}).  In practice, it is convenient to
solve the entire problem with a standard ODE solver.  Unfortunately,
there is a wide range of time scales in the problem: the decay time
and the orbital time scales throughout phase space.  In practice,
then, accurate solutions require small time steps.

The additional satellite orbital frequency complicates the spectrum
and, because the orbital radius now oscillates as well as decays,
complicates the representation of the interaction.  However,
qualitatively, the behaviour is similar to the circular case:
low-order resonances such as the ILR provide strong positive torque on
the halo at early times and then become oscillatory.  Unlike the
circular case, an eccentric orbit will deposit its orbital angular
momentum at lower energies because the amplitude of the perturbation
will be larger at smaller radii.  This response of the halo may be a
wake that influences the subsequent evolution of the disk and the
halo.  We will describe our detailed findings for the satellite-halo
interaction for eccentric orbits in a later paper.

\section{Summary}
\label{sec:summary}

This paper shows the \citet[LBK]{Lynden-Bell.Kalnajs:72} secular
torque formula is not a quantitatively accurate description of secular
evolution in galaxies.  The LBK formula assumes that the growth time
of a perturbation driving secular evolution is infinitely long
compared the characteristic dynamical time.  However, not only is the
number of characteristic dynamical times in a galactic age modest, the
growth of a spiral arms, a bar or the decay of a large dwarf satellite
has evolutionary time scales of order 100 Myr to 1 Gyr, uncomfortably
close to the equilibrium galaxy lifetime of 10 Gyr or smaller.  We
have seen that the finite time effects yield a different distribution
of angular momentum in the halo and a different total torque than for
a time-asymptotic system assumed by LBK.  Similarly, it is unlikely
that time-dependent effects will affect time-asymptotic perturbation
theory predictions in planetary systems \cite{Goldreich.Tremaine:79b,
Goldreich.Tremaine:80} where the secular evolution time scale is much
larger than the dynamical time scale.

In \S\ref{sec:method}, I presented generalised the LBK formula for
perturbations of finite duration.  The new formula shares many of the
features of the LBK formula; the main difference is that the Dirac
delta function $\delta(\ldo - m\Omega_p)$ is replaced with a time
integral (see eq. \ref{eq:LBK1}).  Over astronomically-relevant time
scales, the finite-time generalisation shows that transients from the
formation history of the system play an important role in the overall
dynamical evolution.  For example, Figures \ref{fig:pattern} and
\ref{fig:lbk} illustrate the total torque and differential
contributions of the principal resonances during slow down of a bar in
dark matter halo.  The correct result is the net contribution for
approximately 8 resonances (not all shown in Fig. \ref{fig:lbk}) and
therefore these resonances must be accurately represented in an N-body
simulation to obtain the correct result (see Paper 2).

The most important general finding is that the history of galaxy
evolution can not be ignored in understanding and predicting a
particular galaxy's evolution.  These dynamics are illustrated in two
cases: the bar--halo interaction and the satellite--halo interaction.
The details of the bar--halo interaction have been recently described
in Paper 2 and explored in a self-consistent simulation in Paper 3
\citep{Holley-Bockelmann.etal:04}.  Both papers demonstrate that
bar--halo coupling through the ILR agrees with the predictions
presented here.  Both the LBK formula and the time-dependent formula
predict that the ILR resonance dominates the overall angular momentum
transfer. However, the time-dependent evolution spreads the resonant
interaction over a broader range of lower energies that are more
populated in phase space and therefore have a larger overall
evolutionary consequence.  The time-asymptotic ILR is located at
smaller energies and radii and would be inaccessible to most N-body
simulations.  The time-asymptotic approximation is especially poor for
the ILR resonance in a cuspy halo where the time-dependent spread in
frequency corresponds to a large region in radius and energy.  This is
strong encouragement to attempt an understanding of the dynamical
interactions prior to simulation.  One expects a similar halo response
in the satellite--halo interaction because the quadrupole perturbation
from a rotating bar shares much in common with the quadrupole
perturbation from an orbiting satellite.  Indeed, we show that this is
the case.  However, because the symmetries, the radial profile, and
the distribution of frequencies for the two perturbations are
different, the ILR no longer dominates in the satellite interaction.
Nonetheless, the ILR does play a role during the early stages of
orbital evolution and this has been verified in N-body simulation
\citep[see][]{Choi.etal:04}.  The aggregate effect of this mechanism
from a population of substructure will be the subject of a later
investigation.

The wakes excited by interactions between satellites, disks and dark
matter halos give rise to a variety of observable consequences.  For
example, large-scale transport of angular momentum throughout the disk
and to the halo may will tend to decrease the disk scale length.  The
response of the galaxy halo to the surrounding group environment can
propagate features to the inner galaxy, exciting disk waves and
subsequent star formation.  In addition, the interaction between the
satellite and disk will drive bending modes and the energy in these
modes will heat the disk. Detailed simulations for these processes of
disparate scales are difficult.  Time-asymptotic secular perturbation
theory has often been used to estimate evolution and compare with
simulation.  However, we have found that the difference in amplitude
and significance between the time-dependent and time-asymptotic
responses for the bar--halo interaction (Fig. \ref{fig:lbk}) is
dramatic.  If these differences are representative of the difference
in these other processes, our current picture of secular evolution
must be revisited by considering the time-dependent evolution of these
``waves'' in the context of a galaxy's environmental history.

\section*{Acknowledgments}

This work was begun at the Institute for Advanced Study in Princeton
and I thank my host John Bahcall for his hospitality.  Many thanks to
Kelly Holley-Bockelmann for suggestions and a careful reading of this
manuscript.  I would also like to thank Jerry Sellwood for visiting me
at the IAS and for the discussions that motivated a detailed
comparison to the LBK formula.  This work was supported in part by NSF
AST-9802568 and AST-9988146 and by NASA LTSA NAG5-3525.

\label{lastpage}

\end{document}